\documentclass{aa} 
\usepackage[varg]{txfonts}
\usepackage{epsfig,graphicx,lscape,subfigure}
\usepackage{float,longtable,comment}
\usepackage[none,bottom,light,timestamp]{draftcopy}
\usepackage{rotating,amssymb,arydshln}
\usepackage{rotating,amssymb}
\usepackage{natbib}
\newcommand{\kms}{{km\,s}\ensuremath{^{-1}}}
\newcommand{\teff}{\ensuremath{T_\mathrm{eff}}}
\newcommand{\logg}{\ensuremath{\log g}}
\newcommand{\Msun}{\,$\rm{M}_\odot$}
\newcommand{\Lsun}{\,$\rm{L}_\odot$}

\newcommand{\abund}{$\log \epsilon+12$}
\newcommand{\abundN}{$\log \epsilon_N+12$}
\newcommand{\vsini}{\ensuremath{v_{\rm{e}} \sin i}}

\newcommand{\vt}{\ensuremath{v_{\rm{t}}}}

\newcommand{\dvr}{\ensuremath{\Delta v_{\rm{r}}}}
\begin{document}
\title{The VLT-FLAMES Tarantula Survey\thanks{Based on observations collected at the European Organisation for Astronomical Research in the Southern Hemisphere under ESO programme 182.D-0222.} }
\subtitle{XXVII. Physical parameters of B-type main-sequence binary systems in the Tarantula nebula}
\author{R. Garland\inst{1,2}, P. L. Dufton\inst{1}, C. J. Evans\inst{3}, P. A. Crowther\inst{4}, I.D. Howarth\inst{5},   A. de Koter\inst{6}, S. E. de Mink\inst{6}, N.J. Grin\inst{7}, N. Langer\inst{7}, D. J. Lennon\inst{8},  C. M. McEvoy\inst{1,9}, H. Sana\inst{10}, F.R.N. Schneider\inst{11},  S. S\'imon D\'iaz\inst{12,13}, W. D. Taylor\inst{3}, A. Thompson\inst{1}, J. S. Vink\inst{14}}
{\institute{Astrophysics Research Centre, School of Mathematics and Physics, Queen's University Belfast, Belfast BT7 1NN, UK
        \and{Sub-department of Atmospheric, Oceanic and Planetary Physics, Department of Physics, University of Oxford, Oxford, OX1 3RH, UK}
	\and {UK Astronomy Technology Centre, Royal Observatory Edinburgh, Blackford Hill, Edinburgh, EH9 3HJ, UK}
       \and{Department of Physics and Astronomy, Hounsfield Road, University of Sheffield, S3 7RH, UK}
        \and{Department of Physics and Astronomy, University College London, Gower Street, London WC1E 6BT, UK}
         \and{Anton Pannenkoek Institute for Astronomy, University of Amsterdam, NL-1090 GE Amsterdam, The Netherlands}
         \and{Argelander-Institut f\"{u}r Astronomie der Universit\"{a}t Bonn, Auf dem H\"{u}gel 71, 53121 Bonn, Germany}
	\and{European Space Astronomy Centre (ESAC), Camino bajo del Castillo, s/n Urbanizacion Villafranca del Castillo, Villanueva de la Ca\~{n}ada, E-28692 Madrid, Spain}
	\and{King's College London, Graduate School, Waterloo Bridge Wing, Franklin Wilkins Building, 150 Stamford Street, London SE1 9NH}
       \and{Instituut voor Sterrenkunde, Universiteit Leuven, Celestijnenlaan 200 D, B-3001 Leuven, Belgium}
       \and{Department of Physics, University of Oxford, Keble Road, Oxford OX1 3RH, United Kingdom}
       \and{Instituto de Astrof\'isica de Canarias, E-38200 La Laguna, Tenerife, Spain}              
       \and{Departamento de Astrof\'isica, Universidad de La Laguna, E-38205 La Laguna, Tenerife, Spain}
       \and{Armagh Observatory, College Hill, Armagh, BT61 9DG, Northern Ireland, UK}}
%
%
\date{Received \ Accepted }

\abstract{A spectroscopic analysis has been undertaken for the B-type multiple systems (excluding those with supergiant primaries)  in the VLT-FLAMES Tarantula Survey (VFTS). Projected rotational velocities, \vsini, for the primaries have been estimated using a Fourier Transform technique and confirmed by fitting rotationally broadened profiles. A subset of 33 systems with \vsini$\leq$ 80 \kms\ have been analysed using a TLUSTY grid of model atmospheres to estimate stellar parameters and surface abundances for the primaries. The effects of a potential flux contribution from an unseen secondary have also been considered. For 20 targets it was possible to reliably estimate their effective temperatures (\teff) but for the other 13 objects it was only possible to provide a constraint of 20\,000$\leq$\teff$\leq$26\,000 K -- the other parameters estimated for these targets will be 
consequently less reliable. The estimated stellar properties are compared with evolutionary models and are generally consistent with their membership of 30 Doradus, while the nature of the secondaries of 3 SB2 system is discussed. A comparison with a sample of single stars with \vsini$\leq$ 80 \kms\ obtained from the VFTS and analysed with the same techniques implies that the atmospheric parameters and nitrogen abundances of the two samples are similar.  However, the binary sample may have a lack of primaries with significant nitrogen enhancements, which would be consistent with them having low rotational velocities and having effectively evolved as single stars without significant rotational mixing. This result, which may be actually a consequence of the limitations of the pathfinder investigation presented in this paper, should be considered as a motivation for spectroscopic abundance analysis of large samples of binary stars, with high quality observational data.
}

\authorrunning{R. Garland et al.}

\titlerunning{B-type Main-Sequence Binaries in the Tarantula Nebula}

\maketitle
\nopagebreak[4]

%
\section{Introduction}                                         \label{s_intro}

The quantitative analysis of the spectra of early-type stars provides
a powerful tool for understanding their formation and evolution. With
the development of both observational and theoretical techniques is
has been possible to move from the pioneering analyses of single
bright targets in our Galaxy (see, for example, \citealt{uns42}) to
large-scale surveys in external galaxies \cite[see, for
  example,][]{eva08, eva11}. Currently quantitative spectroscopy is
possible to a distance of several megaparsecs (e.g. \citealt{kur16}),
with the next generation of ground-based telescopes expected to reach
beyond 10~Mpc \citep{eva11a}.

Complementing these observational advances has been an improved
understanding of how early-type stars evolve \citep[see, for
  example,][]{mae09, lan12}. This includes the importance of rotation
in both the evolution of the star \citep{heg00a, hir04} and in the
mixing of nucleosynthesised material to the stellar surface
\citep{mae87, heg00b, fri10}.  Additionally, mass loss can be significant,
particularly for more luminous stars and in metal-rich
environments \citep{pul08,mok07} and this can effect their evolution \citep{chi86}.  
More recently it has been recognised that
magnetic fields \citep{don09} can affect the internal structure and rotation profile, 
which in turn can affect chemical mixing. For example, \citet{pet15} discuss the
effects of magnetic fields on the mixing between  a stellar core and its envelope.
Additionally stellar oscillation modes or 
internal gravity waves \citep[][]{aer14a}  
could be important. These developments have allowed the generation of
extensive grids of evolutionary models \citep[see, for
 example,][]{bro11a, eka12, geo13}, which have been used to synthesize
stellar populations for comparison with the results from large-scale
surveys \citep{bro11b, gri16}.

Most quantitative analyses of stellar spectroscopy have implicitly or
explicitly assumed that their targets have evolved as single
stars. However, it has recently become apparent that most early-type
stars are in binary systems \citep{mas09}. For example, \citet{san12}
estimated an intrinsic binary fraction of 0.69$\pm$0.09 for Galactic O-type
systems, with a strong preference for closely bound
systems. Other Galactic studies include \citet[][binary fraction with periods less than 5000 days of approximately 0.55]{kob14}, \citet[][lower limit of 0.17 for O-type systems]{mah09},  \citet[][0-0.33 for O-type systems]{mah13} and \citet[][$0.30^{+0.34}_{-0.21}$ for O-type systems]{pfu14}. 
In the Large Magellanic Cloud (LMC),  \citet{san13} estimated a fraction of
0.51$\pm$0.04 for the O-type stellar population of the 30~Doradus
regio observed in the VLT-FLAMES Tarantula survey
\citep[VFTS;][hereinafter Paper~I]{eva11}. Additionally \citet{dun15}
inferred a similar high fraction (0.58$\pm$0.11) for the B-type
stellar population in the same survey.

The above estimates are for the current intrinsic fraction of binary
systems. Theoretical simulations \citep{deM14} imply that a
significant fraction  of massive stellar
systems (typically $\sim$8\%) might be the products of stellar mergers. In turn this would
imply an even higher fraction ($\sim$19\%) of {\em currently} single
stars have previously been in a binary system. For example,
\citet{deM14} discuss the possibility that the higher fraction of
binaries found in young Galactic O-type stars \citep{san12} than in
the VFTS sample \citep{san13} might be due to stellar evolution and
binary interactions having modified the intrinsic binary fraction of
the latter. Although there remains some uncertainty in the intrinsic
binary fractions (as a function of age and metallicity) and in the
relative importance of the diverse evolutionary channels available to
binary systems, it is clear that binarity must play a central role in
the evolution of early-type stars.

Given the above, it may appear surprising that there have been few
quantitative analyses of the spectra of early-type binaries (see
Sect.~\ref{s_N_d}). This probably reflects the difficulty of
identifying and observing binary systems and also the additional
complexity of allowing for the flux contributions of the
secondaries. Although methods exist for disentangling the components in
high-quality spectroscopy of SB2 systems \citep[see, for
  example,][]{gon06,how15}, these are not readily applicable to SB1 systems or to
spectroscopy with moderate signal-to-noise ratios.

Here we present what we believe to be the first model-atmosphere analysis of a significant sample of B-type spectroscopic binaries that are not supergiants. These were taken from the binary sample identified in the VFTS \citep{dun15} but were limited to those stars with relatively narrow metal-absorption lines in order to facilitate their
analysis. We emphasize that, given the complexities of
modelling the spectra of binary systems and the biases in our
sample, this should be considered as a pathfinder analysis, although
we believe that the scientific results will still be useful in
constraining theoretical models.

The observational data are discussed in Sect.~\ref{s_obs} and the
analysis in Sect.~\ref{s_parameters}.  Results are
discussed in Sect.~\ref{s_discussion}, which contains a comparison
with a sample of apparently single stars from the same
survey.

\section{Observations}\label{s_obs}

As described in Paper~I and \citet{mce14}, the Medusa mode of FLAMES \citep{pas02} was used to collect the VFTS data. This uses fibres to observe up to 130 sky positions simultaneously with the Giraffe spectrograph.  FLAMES has a corrected field-of-view of 25\arcmin and hence with one telescope pointing, we were able to observe stars  both in the local environs of 30 Doradus as well as in the main body of the H II region. Nine Medusa fibre configurations (Fields~`A' to `I' with an identical field centre) were employed to obtain our sample of approximately 800 stellar objects. Two standard Giraffe settings were used, viz.  LR02 (wavelength range from 3960 to 4564\AA\ at a spectral resolving power of R$\sim$7\,000) and LR03 (4499-5071\AA, R$\sim$8\,500). Details of target selection, observations, and data reduction have been given in Paper~I.

\subsection{Sample selection} \label{s_sample}

The B-type candidates identified in Paper~I and subsequently classified by \citet{eva15} were analysed by \citet{dun15} using a cross-correlation technique. Eleven supergiant and 90 lower-luminosity targets with radial-velocity variations that were both statistically significant and had amplitudes larger than 16~\kms\ were classified as binary.\footnote{In principle, some of these systems could contain more than two stars.}
A further 17 supergiants and 23 lower-luminosity targets were classified as `RV variables' by  \citet{dun15}; we exclude those targets as their radial-velocity variations may not be due to binarity.  

The supergiants have been discussed by \citet{mce14} and will not be considered further here.  Projected rotational velocities (\vsini) have been estimated for the primaries of the remaining 90 binary candidates, using similar techniques to those adopted by \citet{duf12}. Details of the methodology are provided in Appendix~\ref{vsini}, whilst the estimates are given in Tables~2 and~3 (only available online). These have a similar format to the online Tables~3
and~4 in \citet{duf12}.

A quantitative analysis of all these binaries including those with large projected rotational velocities would be possible in principle. However, because of the moderate signal-to-noise (S/N) ratios in our spectroscopy, the atmospheric parameters would be poorly constrained (in particular, effective temperatures -- see Sect.\ \ref{s_Teff}). In turn this would lead to nitrogen abundance estimates that had little diagnostic value. Hence for the purposes of this paper we have limited our sample to the subset of binaries with relatively small projected rotational velocities.

The thirty-seven binaries with projected equatorial rotational velocities $\vsini \leq 80$~\kms\ are listed in Table~\ref{t_Targets}, along with their spectral types \citep{eva15, wal14} and typical S/N ratios and observed range of radial velocity variations, \dvr\ \citep{eva15}. As the LR02 wavelength region 4200--4250\AA~should not contain strong spectral lines, our S/N ratios were estimated from that region. 
We should consider these as lower limits, however, as weak absorptions lines could have an affect (especially for higher S/N ratios). S/N ratios for the LR03 region were normally similar to or slightly smaller than those for the LR02 region \citep[see, for example,][for a comparison of S/N ratios in the two spectral regions]{mce14}.  The estimates of \dvr \ will be related to the amplitude of the primary's radial velocity variations (depending on the sampling of the orbit they will be similar to but generally lower than twice the amplitude) and hence provide insight into whether the systems are tightly bound.

Four targets proved resistant to quantitative analysis (see Sect.\ \ref{s_meth}), leaving 33 for which a detailed study could be undertaken. For the most part, these are single-lined (SB1) systems as far as our
data are concerned;  only three targets, discussed in
Sect.~\ref{s_SB2}, show any direct evidence for a secondary spectrum.

\begin{table}
  \caption{Spectral classifications \citep[from][]{eva15}, estimated
    projected rotational velocity (\vsini), typical signal-to-noise
    (S/N) ratios for the LR02 region (which have been rounded to the
    nearest multiple of five) and observed ranges of radial velocity variations of the primary, \dvr. The SB2 or SB2? classifications have been updated on the basis of  the discussion in Sect.\ \ref{s_SB2}. Also listed are the estimates of the
    luminosities discussed in Sect.\ \ref{s_HR}, apart from the 4 stars that
    were not analysed using model atmosphere techniques.}\label{t_Targets} \centering
\begin{tabular}{clrrrc}
\hline\hline
Star &Spectral type & \vsini & S/N & \dvr & $\log$ L/\Lsun \\
\hline
017 &                 	B0 V  		        & 76           &  90  & 72  & 4.85 \\
018 &               	B1.5 V  		        & 48           & 30   & 21  & 4.20\\
033 &             		B1--1.5 V  		& 77           &  85  & 51  & 4.26\\
041 &                    	B2: V 		        & $\leq$40 & 35   & 40  & 3.95\\
097 &         		B0 IV  		        & 72           & 65   & 24  & $\cdots$ \\
162 &               	B0.7 V  		        & 60           & 60   & 32  & 4.22\\
179 &                 	B1 V  		       & 51            & 50    & 17  & 4.03 \\
195 &               	B0.5 V  		       & $\leq$40  & 60    & 29  & 4.06\\
204 &               	B2 III  		       & $\leq$40  & 70    & 31  & 4.41\\
218 &               	B1.5 V  		       & 79            & 85    & 20  & 4.89\\
225 &                       B0.7--1 III--II          & $\leq$40  & 95     & 32  & 4.53\\
240 &         		B1--2 V (SB2)         & 77           & 100  & 152 & $\cdots$\\
278 &               	B2.5 V  		       & 60           &100   &  33  & $\cdots$\\
299 &               	B0.5 V  		       & $\leq$40  & 70   &  108 & 4.14\\
305 &                	B2: V  		       & 57            & 60   &  123 & $\cdots$ \\
324 &               	B0.2 V  		       & 57            &105  &  64   & 4.42\\
342 &                 	B1 V  		       & $\leq$40  & 55   &  104 & 3.84\\
351 &               	B0.5 V  		       & $\leq$40  & 90   &  58   & 4.47\\
359 &               	B0.5 V  		       & 54            & 70   &  26   & 4.27\\
434 &              		B1.5: V  		       & 45            & 85   &  54   & 4.38\\
501 &               	B0.5 V  		       & 59            & 140 &  141 & 4.57 \\
520 &         		B1: V (SB2?)	       & 53            & 75   &  180 & 4.11\\
534 &            		B0 IV  	               & 57            & 55   &  59   & 4.82\\
575 &             		B0.7 III 		       & $\leq$40  & 100 &  56    & 4.57\\
589 &         		B0.5 V  (SB2)         & $\leq$40  & 60   & 179   & 4.63  \\
662 &                       B3-5 III:                  & 67            & 90   & 100   & 3.94\\
665 &               	B0.5 V  		       & 47            & 65   &  37    & 4.33\\
686 &             		B0.7 III (SB2)	       & $\leq$40  & 125 &  90    & 4.83 \\
719 &                 	B1 V  		       & 50            & 50   &  52    & 3.99 \\
723 &               	B0.5 V  		       & 63            & 75   &  62    & 4.50\\
742 &                	B2 V  		       & 60            & 40   &  21    & 3.73 \\
792 &                 	B2 V  		       & 47            & 75   &  18    & 4.07 \\
799 &          	 	B0.5--0.7 V  	       & $\leq$40  & 35   & 20     & 4.13 \\
850 &               	B1 III  		       & $\leq$40  & 40   & 27     & 4.34 \\
874 &            		B1.5 IIIe+  	       & 62            & 90   &  20    & 4.37\\
888 &               	B0.5 V  		       & 76            & 75   & 121   & 4.18 \\
891 &                 	B2 V  		       & 55            & 50   & 21     & 4.04 \\
\hline
\end{tabular}
\label{tab.basics}
\end{table}

\subsection{Data preparation}

Data reduction followed the procedures discussed in \citet{eva11} and
\citet{duf12}. Because of the radial-velocity variations between
epochs, care had to be taken when combining exposures. We undertook
numerical experiments (discussed in Appendix~\ref{vsini}) to estimate
the maximum range (\dvr) in radial-velocity measurements before the
estimation of the projected rotational velocity became
compromised. For slowly rotating targets, this was found to occur at
$\dvr \simeq 30$~\kms, and this should also be an appropriate range
for a model-atmosphere analysis. Hence for targets with $\dvr \leq
30$~\kms\, we combined all usable exposures for the LR02 spectra, without applying any radial-velocity shifts, using either a median or weighted $\sigma$-clipping algorithm. The two methods gave final
spectra that were effectively indistinguishable. Using only exposures from a single epoch (with effectively constant radial velocity), gave comparable but lower S/N ratio results.

For those stars with greater values of \dvr\, most of the treatment was the same, except that either ($i$) spectra were shifted to allow for
radial-velocity variations prior to being combined, and/or ($ii$)
spectra from the highest S/N ratio LR02 epoch and
additional epochs with radial velocities within $\pm$15~\kms\ were combined.  The latter
procedure minimised the effects of nebular contamination, whilst the
former led to higher S/N ratios.

The time cadence for the LR03 exposures are listed in Tables A.1 and A.2 in the Appendix of Paper I. 
For five of the Medusa fields all the exposures were taken within a
period of 3 hours, whilst for two other fields the spectroscopy
was obtained on consecutive nights. For Field~F, there was an
additional observation separated by 55~days from the other six exposures
but this exposure was not included in the reduction. Hence for 
these eight fields, simply combining exposures should be adequate,
especially as no significant radial-velocity shifts (i.e., greater than
30~\kms) were observed in LR02 exposures taken on the same or
consecutive nights.

For Field~I (containing eight targets: VFTS~017, 018, 278, 534, 575, 742,
850, 874), one set of LR03 exposures was separated from the others
by 8 days.  One target  was not analysed (VFTS\,278;  see
Sect.~\ref{s_meth}) but for the other seven targets the individual
co-added spectra for the two epochs were cross-correlated. The
wavelength region 4530--4720\AA\ was selected, as this contains
relatively strong metal and helium absorption lines. The
cross-correlations implied velocity shifts of less than 30~\kms\ and
hence the individual exposures for these targets were again combined
without any wavelength shifts.

\section{Model-atmosphere analyses} \label{s_parameters}

\subsection{Methodology}\label{s_meth}

We employed model-atmosphere grids calculated with the {\sc tlusty}
and {\sc synspec} codes \citep{hub88, hub95, hub98, lan07}.  
They cover a range of effective temperature, 10\,000K $\leq$\teff$\leq$35\,000K in steps of typically 1\,500K.  Logarithmic gravities (in cm s$^{-2}$)  range from 4.5 dex down to the Eddington limit in steps of 0.25 dex, and microturbulences are from 0-30 \kms in steps of 5 \kms. As discussed in  \citet{rya03} and \citet{duf05}, equivalent widths and line profiles interpolated within these grids are in good agreement with those calculated explicitly at the relevant atmospheric parameters.

These
codes adopt the `classical' non-LTE assumptions, i.e. plane-parallel atmospheres and that the optical spectrum is unaffected by winds. As the targets
considered here have luminosity classes III--V, such an approach
should provide reliable results. These models assume a normal helium to hydrogen ratio (0.1 by number of atoms). The validity of this will be considered in Sect.\ \ref{GC}. Grids have been calculated for a
range of metallicities (see \citealt{rya03, duf05} for details) with
that for an LMC metallicity being used here.

For four targets  (VFTS\ 097, 240, 278, 305), the \ion{Si}{iii}
spectra could not be reliably measured, thereby precluding
estimation of the microturbulence (and also, when the \ion{He}{ii} spectrum
was absent, the effective temperature); these stars were excluded
from the analysis. Another 13 targets had no observable
\ion{Si}{ii},  \ion{Si}{iv} or \ion{He}{ii} features in their spectra,
leading to uncertainties in their effective-temperature estimates of
more than $\pm$2000\,K (see Sect.\ \ref{s_Teff} for details of
methodology). These were initially excluded from our analysis on the
basis that their atmospheric parameters (and hence nitrogen
abundances) would be unreliable. However, the \ion{N}{ii} lines have a
maximum strength at the atmospheric parameters appropriate to these
stars.  This leads to the nitrogen abundance  being relatively
insensitive to the choice of atmospheric parameters.  Hence these
stars have been analysed using a modified methodology as discussed in Sect.\ \ref{s_Other}. 

For our targets, there will be an indeterminate flux contribution
from the fainter secondary. To investigate the possible effects of this on our
analysis, we assume that the secondary contributes 20\%\ of the light
via a featureless continuum, just as in \citet{mce14}. We have chosen this value as contributions larger than this should become apparent in the secondary spectrum
(but see the discussion in Sect.\ \ref{s_SB2}). A simple continuum was
adopted for convenience and also because structure in the secondary
spectrum (e.g. in the Balmer lines) would normally mitigate its
effect. This analysis used the same methodology to that discussed
below apart from the equivalent widths and hydrogen and helium line
profiles having been rescaled; this approach was used previously
by \citet{mce14, dun11}, where more details can be found.

Due to the interdependence of the estimation of the effective temperature, surface 
gravity, microturbulence, and chemical abundances, an iterative process
was adopted as discussed below.

\subsection{Effective Temperature} \label{s_Teff}

Effective temperatures (\teff) were initially constrained by the
silicon ionisation balance, using microturbulence estimates from the
absolute silicon abundance (method 2 discussed in Sect.\ \ref{s_vt}).
The equivalent widths of the \ion{Si}{iii} triplet (4552, 4567, 4574\AA) were
used in conjunction with those of either a \ion{Si}{iv} line (4116\AA) for the
hotter systems, or a \ion{Si}{ii} doublet (4128, 4130 \AA) for the
cooler ones. For some systems, neither the \ion{Si}{ii} nor the
\ion{Si}{iv} lines were observable; in these cases, upper limits were
set on their equivalent widths, allowing limits for the effective
temperature to be estimated. Stars, where the uncertainty in the  $\teff$ estimates was greater 
than $\pm$2000\,K, were excluded from this analysis
but are discussed further in Sect.~\ref {s_Other}.

Where \ion{He}{ii} lines were present in the spectrum, two independent
effective temperature estimates were made from fitting the line
profiles of the features at 4541 and 4686\AA \ (with the theoretical profiles being convolved with the instrumental profile and with a rotational broadening function), assuming gravities estimated from the hydrogen line profiles (see Sect.\ \ref{m_g})  and a normal helium to hydrogen abundance. The latter assumption is considered further in Sect.\ \ref{GC}. All the effective
temperature estimates are summarised in Table~\ref{t_Single_2} for the
assumption that the secondary contributed no flux; the estimates
assuming that the secondary contributed 20\%\ of the flux were
generally higher by 500--1000\,K, as can be seen from
Table~\ref{t_Atm}.

Generally the estimates from the silicon and helium spectra were in
good agreement, with the difference being 1\,000\,K or less (see
Table~\ref{t_Single_2}). This is consistent with the S/N ratios of our spectroscopy 
(see Table \ref{t_Targets}), which lead to formal uncertainties in our equivalent widths estimates 
of typically less than 10\%. We have therefore adopted a
 stochastic uncertainty in our effective-temperature estimates
of $\pm$1\,000\,K.

\begin{table}

\caption{Estimates of the atmospheric parameters for the primaries
  assuming no secondary flux contribution. Effective-temperature
  estimates are from the silicon ionisation balance (Si) or the
  \ion{He}{ii} profiles at 4541 and 4686\AA. The former used the
  microturbulent velocity estimates from the absolute silicon
  abundance (method 2). Microturbulence estimates are from the
  relative strengths of the \ion{Si}{iii} triplet (1) or the absolute
  silicon abundance (2).}\label{t_Single_2}

\begin{center}
\begin{tabular}{lcccccc}
\hline\hline
\vspace*{-0.25cm}\\
Star  & \multicolumn{3}{c}\teff & \logg & \multicolumn{2}{c}\vt \\
& Si & $\lambda$4542 & $\lambda$4686 & & (1) & (2)\\
\hline
017 & 29000 & 30000 & 29500 & 3.90 & 6 & 2 \\
033 & \,\,24000$^\dagger$ & $\cdots$ & $\cdots$ & 3.90 & 3 & 1 \\
179 & 27000 & $\cdots$ & $\cdots$ & 4.40 & 0 & 0 \\
195 & 28000 & $\cdots$ & 28000 & 3.90 & 0 & 0 \\
204 & \,\,22500$^\dagger$ & $\cdots$ & $\cdots$ & 3.50 & 0 & 0 \\
225 & 24500 &  $\cdots$ & $\cdots$  &	3.25  & 3 & 5\\
299 & 28000 & $\cdots$ & 29000 & 4.25 & 0 & 0 \\
324 & 28500 & 29000 & 29000 & 3.90 & 0 & 0 \\
351 & 28500 & $\cdots$ & 28500 & 4.00 & 0 & 0 \\
359 & 28000 & 28500 & 28500 & 4.00 & 0 & 1 \\
534 & 29000 & 29000 & 30000 & 3.75 & 13 & 5 \\
575 & 26000 & $\cdots$ & 25000 & 3.75 & 0 & 4 \\
589 & 27500 & $\cdots$ & 28000 & 4.00 & 0 & 0 \\
662 & 17500 & $\cdots$ & $\cdots$ & 3.60 & $\cdots$ & 7 \\
665 & 28000 & $\cdots$ & 27500 & 4.15 & 1 & 2 \\
686 & 24000 & $\cdots$ & $\cdots$ & 3.60 & 4 & 3 \\
723 & 27500 & $\cdots$ & 27000 & 3.90 & 0 & 2 \\
799 & 26500 & $\cdots$ & 26000 & 4.00 & 2 & 1 \\
850 & 24000 & $\cdots$ & $\cdots$ & 3.75 & 22 & 6 \\
888 & 27000 & $\cdots$ & 27000 & 4.15 & 6 & 0 \\
\hline
\end{tabular}
\end{center}
~\\
$^\dagger$: \teff~estimates from absence of  \ion{Si}{ii} and \ion{Si}{iv} lines.

\end{table}

\subsection{Surface Gravity}\label{m_g}

The observed hydrogen Balmer lines profiles, H$\gamma$ and H$\delta$, were compared to theoretical profiles (again convolved to allow for the instrumental profile and stellar rotation) in order to estimate the logarithmic surface gravities (\logg; cm s$^{-2}$); further details can be found in, for
example, \citet{mce14}. These estimates are summarised in
Table~\ref{t_Single_2} for the assumption that the secondary
contributed no flux; the estimates assuming that the secondary
contributed 20\%\ of the flux are larger by between 0.1--0.3~dex
(mainly due to the observed hydrogen line profiles being deeper) as
can be seen from Table~\ref{t_Atm}.

Estimates derived from the H$\gamma$ and H$\delta$ lines generally
agreed to $\pm$0.1~dex. Factoring in other uncertainties, such as
normalisation errors and possible errors in the line broadening
theory, a conservative error estimate of $\pm$0.2~dex has been adopted
(note that the estimates are quoted in the Tables to the nearest
0.05~dex, i.e. two significant figures, in order to illustrate the
effects of the secondary flux contribution).

\subsection{Microturbulence} \label{s_vt}

By eliminating any systematic dependence of abundance estimates on line strength, we can estimate the microturbulent velocities (\vt). To do this, we used  the equivalent widths of
the \ion{Si}{iii} triplet (4552, 4567 and 4574\AA) because it is present in almost all of our spectra.  All three lines are from one multiplet, which has the advantages that their {\em relative} oscillator strengths should be reliable, whilst non-LTE effects should be similar for each line. The only other choice for estimating the microturbulence, given our wavelength coverage, would have been to use the rich \ion{O}{ii} spectra. However as discussed by \citet{sim06} and \citet{sim10a}, this approach has several complications. Firstly many of the lines are blended, leading \citet{sim06} to reject over half the features they identified. Secondly, errors in the adopted oscillator strengths or in the non-LTE effects for different multiplets can lead to systematic errors. A third difficulty specific to our observational dataset was that the \ion{O}{ii} spectrum was weak in our coolest targets and was not therefore a sensitive diagnostic for the microturbulence.

\begin{figure}
\epsfig{file=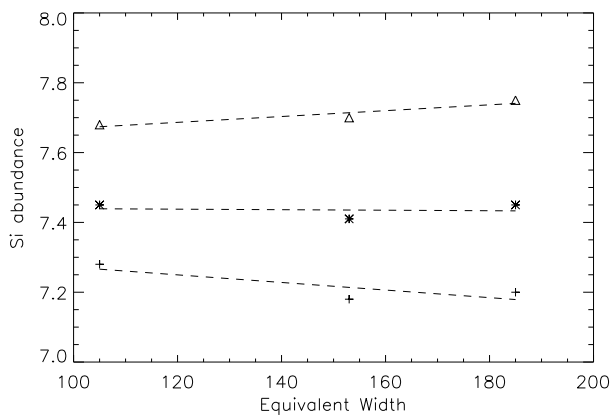,width=\linewidth, angle=0} 
\caption{An example of the estimation of the microturbulent velocity using the relative strengths of three \ion{Si}{iii} lines in VFTS\,225. Abundances estimates are shown for \vt =1 (triangles), 3 (stars) and 5 (crosses) \kms, together with linear least square fits.}
\label{vt}
\end{figure}

 Our methodology  is illustrated in Fig. \ref{vt}  for VFTS\,225. As can be seen the slope of the least squares fit of the abundance estimates against equivalent widths is relatively insensitive to the choice of microturbulence. Therefore errors in the equivalent widths estimates can lead to significant errors in the estimation of the microturbulence, especially when the lines lie close to the linear part of the curve of growth \citep[see,][for more details]{mce14, duf05, hun07}. The method also requires that all three lines be observed reliably, which was not always the case with the current dataset. Additionally, large microturbulence values were obtained for VFTS\,534 and VFTS\,850, which appeared inconsistent with their derived gravities. 

Therefore the microturbulence estimates were also obtained by ensuring that the silicon abundance is
consistent with the LMC's metallicity,
since this element should not be significantly affected by nucleosynthesis in our targets \citep[see, for example,][]{bro11a}.
A value of 7.20~dex was adopted as found by \citet{hun07} using similar
analysis methods and observational data to those conducted here. Previous investigations of early-type stars in the LMC have found similar values, with for example \citet{kor02, kor05} finding respectively 7.10$\pm$0.07 dex in 4 narrow lined (near) main sequence targets and 7.07 dex (with uncertainties in the individual measurements of $\pm$0.3 dex) for 3 broad lined targets. 

To investigate the sensitivity of our estimates to the adopted silicon abundance, we considered the effect of reducing our adopted silicon abundance to 7.1 dex. This led to a typical increase of 1 \kms with a maximum increase of less than 2 \kms. Additionally Fig. \ref{vt} confirms that {\em absolute} silicon abundance estimates are sensitive to the choice of the microturbulence (the silicon abundance estimate changing from approximately 7.7 dex to 7.2 dex as the microturbulence is increased from 1 to 5 \kms), making this potentially a reliable method for estimating this quantity.

Estimates using the two methodologies are summarised in Table~\ref{t_Single_2} for the
assumption that the secondary contributed no flux; the estimates
assuming that the secondary contributed 20\%\ of the flux are generally
larger by 2--5~\kms\ as can be seen from Table~\ref{t_Atm}.

Some systems were found to have a maximum silicon abundance (obtained
at $\vt = 0$~\kms) below the adopted LMC value; furthermore, removing the variation of the abundance estimates with line strength was not always possible.
These effects were mitigated by allowing a flux
contribution from a secondary (see Table~\ref{t_Atm}). However, both
\citet{mce14} and \citet{hun07} found similar effects for a number of
their (presumably single) targets, with the latter discussing possible
causes in some detail. In these instances, we have assumed the
best-estimate microturbulence (i.e., zero).

For other targets we adopted an equivalent-width uncertainty of
$\pm$20\%, normally translating to variations of up to $\sim$5~\kms\
for both methods.  Such uncertainties are consistent with the
differences between the estimates using the two methodologies. Only in
three cases do these differ by more than 5~\kms\ and in two of these
the estimates from relative strength of the \ion{Si}{iii} multiplet
(method~1 in Table~\ref{t_Single_2}) appear inconsistent with the
relatively high surface-gravity estimates.  Hence a reasonable estimate
for the uncertainty in the microturbulence estimates is
$\pm$5~\kms.

\subsection{Adopted atmospheric parameters}\label{Ad_ap}

Table~\ref{t_Atm} summarises the adopted atmospheric parameters. The
parameters are provided for the two different assumptions concerning
the secondary flux contribution, viz. the primary star supplies all
the observed radiative flux or that 20\%\ of the continuum flux is
from the secondary. Microturbulence estimates were from the
requirement of a normal silicon abundance (method 2 in
Table~\ref{t_Single_2}).

Magnesium abundances could be independently estimated from the
equivalent width of the \ion{Mg}{ii} doublet at 4481\AA. This element should not be affected
by nucleosynthesis in (near) main-sequence stars. Hence it
could be used to validate our adopted atmospheric parameters by
comparing estimated abundances to the baseline LMC value
\citep[7.05~dex, as found by][]{hun07}.

The mean (and standard deviations) of these magnesium abundance
estimates was 6.95$\pm$0.15~dex (no secondary flux contribution) and
7.08$\pm$0.15~dex (20\%\ secondary flux contribution). These are both
in reasonable agreement with the baseline LMC abundance and indeed are
consistent with the secondary flux contributions being within the
range considered.

\begin{table*}

\caption{Adopted estimates for the projected rotational velocity
  (\vsini), atmospheric parameters, silicon, magnesium and
  nitrogen abundance estimates. For convenience the spectral types are
  repeated from Table~\ref{t_Targets}.} \label{t_Atm}

\begin{center}
\begin{tabular}{llrllllrrr}
\hline\hline
Star & Spectral Type  & \vsini & Secondary   & \teff  &  \logg & \vt & \multicolumn{3}{c}{\abund}  \\
	&  & \kms\ & Contribution &	K & cm s$^{-2}$	& \kms\	& Si & Mg & N 
 \\	\hline
017 & B0 V & 76 & 0\%\ & 29000 & 3.90 & 2 & 7.22 & 7.11 & $\leq$7.1 \\
 &  &  & 20\%\ & 30000 & 4.15 & 7 & 7.19 & 7.21 & $\leq$7.3 \\
033 & B1--1.5 V  & 77 & 0\%\ & 24000 & 3.90 & 1 & 7.21 & 7.11 & $\leq$6.7 \\
 &  &  & 20\%\ & 23500 & 4.15 & 6 & 7.21 & 6.95 & $\leq$6.8 \\
179 &                       B1 V & 51 & 0\%\ & 27000 & 4.40 & 0 & 6.90 & 7.16 & $\leq$7.2 \\
 &  &  & 20\%\ & 27500 & 4.50 & 0 & 7.21 & 7.38 & $\leq$7.4 \\
195 &                     B0.5 V & $\leq$40 & 0\%\ & 28000 & 3.90 & 0 & 7.07 & 6.86 & $\leq$7.1 \\
 &  &  & 20\%\ & 28000 & 4.00 & 2 & 7.23 & 7.00 & $\leq$7.2 \\
204 & B2 III  & $\leq$40 & 0\%\ & 22500 & 3.50 & 0 & 6.99 & 7.02 & $\leq$6.7 \\
 &  &  & 20\%\ & 23000 & 3.75 & 0 & 7.03 & 7.26 & $\leq$6.9 \\
225 & B0.7 III--II  & $\leq$40 & 0\%\ & 24500 & 3.25 & 5 & 7.22 & 7.03 & $\leq$7.1  \\
 &  &  & 20\%\ &  25000& 3.45 & 8 & 7.19 & 7.16 & $\leq$7.3 \\
299 &                     B0.5 V & $\leq$40 & 0\%\ & 28000 & 4.25 & 0 & 7.14 & 6.83 & $\leq$7.3 \\
 &  &  & 20\%\ & 29000 & 4.50 & 4 & 7.20 & 6.99 & $\leq$7.5 \\
324 &                     B0.2 V & 57 & 0\%\ & 28500 & 3.90 & 0 & 7.19 & 7.12 & $\leq$7.0 \\
 &  &  & 20\%\ & 30000 & 4.25 & 5 & 7.18 & 7.23 & $\leq$7.1 \\
351 &                     B0.5 V & $\leq$40 & 0\%\ & 28500 & 4.00 & 0 & 7.09 & 7.00 & $\leq$7.0 \\
 &  &  & 20\%\ & 29000 & 4.25 & 3 & 7.22 & 7.12 & $\leq$7.2 \\
359 &                     B0.5 V & 54 & 0\%\ & 28000 & 4.00 & 1 & 7.17 & 7.07 & $\leq$7.1 \\
 &  &  & 20\%\ & 29500 & 4.40 & 6 & 7.22 & 7.18 & $\leq$7.2 \\
534 &            B0 IV & 57 & 0\%\ & 29000 & 3.75 & 5 & 7.20 & 6.86 & $\leq$7.2 \\
 &  &  & 20\%\ & 29500 & 4.00 & 9 & 7.23 & 7.08 & $\leq$7.2 \\
575 &                   B0.7 III & $\leq$40 & 0\%\ & 26000 & 3.75 & 4 & 7.24 & 7.02 & 6.90 \\
 &  &  & 20\%\ & 26500 & 4.00 & 8 & 7.21 & 7.13 & 7.03 \\
589 &                     B0.5 V & $\leq$40 & 0\%\ & 27500 & 4.00 & 0 & 6.84 & 6.75 & $\leq$6.9 \\
 &  &  & 20\%\ & 28500 & 4.25 & 0 & 7.14 & 6.92 & $\leq$7.1 \\
662 &            B3--5 III:  & 67 & 0\%\ & 17500 & 3.60 & 7 & 7.16 & 6.80 & $\leq$7.4 \\
 &  &  & 20\%\ & 18000 & 3.90 & 11 & 7.22 & 6.94 & $\leq$7.5 \\
665 &                     B0.5 V & 47 & 0\%\ & 28000 & 4.15 & 2 & 7.16 & 7.00 & $\leq$7.0 \\
 &  &  & 20\%\ & 28500 & 4.40 & 7 & 7.17 & 7.12 & $\leq$7.2 \\
686 &                    B0.7 III & $\leq$40 & 0\%\ & 24000 & 3.60 & 3 & 7.16 & 6.53 & 6.65 \\
 &  &  & 20\%\ & 25000 & 3.90 & 6 & 7.20 & 6.68 & 6.76 \\
723 &                     B0.5 V & 63 & 0\%\ & 27500 & 3.90 & 2 & 7.19 & 6.93 & 7.02 \\
 &  &  & 20\%\ & 28500 & 4.15 & 6 & 7.21 & 7.10 & 7.21 \\
799 &           B0.5--0.7 V  & $\leq$40 & 0\%\ & 26500 & 4.00 & 1 & 7.15 & 6.93 & 7.33 \\
 &  &  & 20\%\ & 27000 & 4.25 & 5 & 7.17 & 7.02 & 7.45 \\
850 &                     B1 III & $\leq$40 & 0\%\ & 24000 & 3.75 & 6 & 7.16 & 6.84 & 7.04 \\
 &  &  & 20\%\ & 26000 & 4.15 & 9 & 7.23 & 7.07 & 7.26 \\
888 & B0.5 V & 76 & 0\%\ & 27000 & 4.15 & 0 & 7.11 & 6.97 & $\leq$7.0 \\
 &  &  & 20\%\ & 28500 & 4.50 & 4 & 7.22 & 7.11 & $\leq$7.2 \\

\hline
\end{tabular}
\end{center}
\end{table*}

\subsection{Nitrogen Abundances} \label{s_N}

The singlet transition at 3995\AA\ is normally the strongest
\ion{N}{ii} line in the LR02 and LR03 spectral regions and appears
unblended \citep[see, for example][]{hun07, mce14}. Therefore estimates 
of its equivalent width together with a curve of growth approach were used 
 as our primary estimator of nitrogen
abundances. Surprisingly this transition was only observable in five
of our systems; by contrast it was observed in approximately
  50\%\ of the equivalent single-star sample analysed
  by Dufton et al. (in prep.). For all of the other B-type binaries, we set an upper
limit on the equivalent width using their methodology.  Briefly, the equivalent widths of weak metal absorption
lines were measured in VFTS spectra with different S/N ratios and
projected rotational velocities. These were then used to infer the
upper limit of the \ion{N}{ii} line in a binary spectrum with a given
S/N ratio and projected rotational velocity.

Three of the five systems where the \ion{N}{ii} 3995\AA\ line was
observed also showed the \ion{N}{ii} line at 4630\AA\ (other weaker
components of this triplet-triplet multiplet could not be
identified). VFTS\,575 and VFTS\,850 yielded nitrogen abundance
estimates which agreed to within 0.1~dex, while VFTS\,723 agreed to
within 0.3~dex.

Estimates and upper limits of the chemical abundances are also affected by the errors in the atmospheric parameters. For further discussion of this, see \citet{hun07} and \citet{fra10}. As in \citet{mce14}, we conservatively adopt a typical uncertainty of 0.2--0.3~dex, which does not include any systematic errors inherent in the adopted models or the secondary flux contribution.

We validated these nitrogen abundance estimates by extracting
theoretical spectra from our grid using the parameters in
Table~\ref{t_Atm}. We have considered both a 0\%\ and 20\%\ contribution from the
secondary, with for the latter the theoretical spectrum being scaled. 
In general there was a good agreement between the
theoretical (convolved to allow for instrumental and rotational
broadening) and observed spectra and Fig.~\ref{N_575} shows some
examples. For two systems (VFTS\,534 and 662), the comparison
indicated that their maximum nitrogen estimates could be up to
0.1--0.2~dex larger than those listed in Table~\ref{t_Atm}.

Figure~\ref{N_575} also shows the \ion{He}{i} line at
4009\AA. Assuming a normal helium abundance, this can be
used as a further check on the atmospheric parameters. For VFTS\,575,
good agreement between theory and observation is found with a small secondary flux contribution, whilst for VFTS\,589 a significant
contribution is implied (this star was classified as SB2 and is
discussed further in Sect.~\ref{s_SB2}).

This consistency check was also carried out for the other eighteen targets and
generally good agreement was found between theory and observation. An attempt was
also made to characterize the amount of secondary contamination and the targets have been divided 
into three groups, viz, those where the contamination appeared small, those where the 
secondary flux contamination appeared to be approximately 20\% and those where the 
contamination lay between 0 and 20\%. For some stars, the comparison 
was complicated by there being evidence of nebular emission in the centre of the
\ion{He}{i}  lines that had not been totally removed by the sky subtraction (due to the sky fibres 
being spatially separated from the target fibres - see Paper I for more details). This was particularly
noticeable for VFTS\,359 and VFTS\,662, which were therefore excluded from this comparison. 
Five stars (VFTS\,033, 299, 324, 665, 888) appeared to have only a small secondary flux contribution,
4 stars (VFTS\,195, 204, 351, 686) had a contribution near the 20\% level and 
7 stars (VFTS\,017, 179, 225, 534, 723, 799, 850) had contributions between 0 and 20\%. 
It is encouraging that VFTS\,686, which had been classified as SB2 showed a significant secondary flux contribution.

The comparison described above was only used as a consistency check and no attempt was made to estimate secondary flux contributions for individual systems. This was because  uncertainties both observational (e.g. possible nebular emission) and in the estimated \vsini\ and atmospheric parameters would have led to significant uncertainties. Rather we have presented nitrogen abundance estimates in Table \ref {t_Atm} for the two representative cases (viz. seconary flux contributions of 0\% and 20\%) that were used to estimate the atmospheric parameters.

\begin{figure}
\epsfig{file=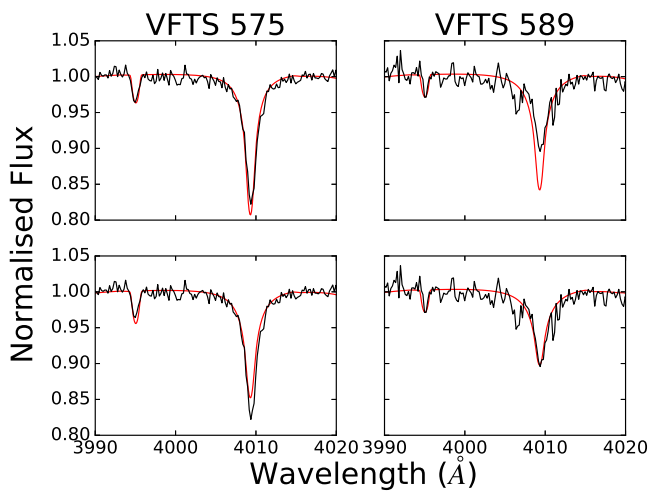,width=\linewidth, angle=0}
\caption{Comparison of observed and theoretical spectra for the region
  near the \ion{N}{ii} line at 3995\AA. VFTS\,575 and VFTS\,589 are
  shown to the left and right respectively. The upper and lower
  panels have a 0\%\ and 20\%\  (with the theoretical spectra being scaled) contribution from the secondary,}
\label{N_575}
\end{figure}

\begin{table}

\caption{Nitrogen abundance estimates and projected rotational
  velocities for targets for which a reliable effective-temperature
  estimate could not be obtained. The gravity estimate is for \teff =
  23\,000 K and a zero microturbulence has been adopted. For
  convenience the spectral types (ST) are repeated from
  Table~\ref{t_Targets}.} \label{t_Oth}

\begin{center}
\begin{tabular}{llccc}
\hline\hline
Star & ST  & \vsini &   \logg & \abundN  \\
	&  & \kms\ &  cm s$^{-2}$	
 \\	\hline
018 &               	B1.5 V  		& 48              &   3.70 & $\leq$7.5 \\
041 &                    	B2: V 		& $\leq$40    &   3.90 & $\leq$7.6 \\
162 &               	B0.7 V  		& 60              &   4.10 & $\leq$7.3 \\
218 &               	B1.5 V  		& 79              &   3.75 & $\leq$6.9 \\
342 &                 	B1 V  		& $\leq$40    &   4.05 & 6.86--7.11 \\
434 &              		B1.5: V  		& 45              &   3.95 & 7.00--7.26 \\
501 &               	B0.5 V  		& 59              &   4.00 & 6.55--6.71 \\
520 &         		B1: V 		& 53              &   4.10 & $\leq$7.3 \\
719 &                 	B1 V  		& 50              &   3.95 & $\leq$7.2 \\
742 &                	B2 V  		& 60              &   4.05 & $\leq$7.4 \\
792 &                 	B2 V  		& 47              &   4.05 & 6.98--7.20 \\
874 &            		B1.5 IIIe+  	& 62              &   3.55 & 6.67--6.81 \\
891 &                 	B2 V  		& 55              &   4.00 & $\leq$7.3 \\
\hline
\end{tabular}
\end{center}
\end{table}

\subsection{Targets with significant uncertainties in \teff}\label{s_Other}

As discussed in Sect.~\ref{s_meth}, there were 13 targets with
observed spectra which had neither observable lines from two
ionization stages of silicon nor observable \ion{He}{ii} profiles. In
these cases, it was not possible to obtain reliable
effective-temperature estimates and they were excluded from the
analysis outlined in Sect.~\ref{s_meth} to \ref{s_N}.

Failure to observe either the \ion{Si}{ii} or \ion{Si}{iv} features
implied that both spectra are relatively weak. This will normally occur
when they have a similar strength. The corresponding effective
temperature will depend on both the adopted gravity and
microturbulence. However, for the typical gravities and
microturbulences found in our sample (see Table \ref{t_Atm}), it is
approximately 23\,000 K.

This effective temperature corresponds to that for the maximum
strength of the \ion{N}{ii} spectra. As the differential of nitrogen
abundance with effective temperature is then zero, this leads to such
estimates being relatively insensitive to the effective
temperature. Additionally for a given observed hydrogen profile, an
increase in the adopted effective temperature leads to an increase in
the estimated gravity (corresponding to approximately 0.1~dex in
\logg\ for a change of 1\,000\,K in \teff). This decreases the
sensitivity of the degree of nitrogen ionization (given in LTE by the
Saha equation) to changes in effective temperature. In turn this
further reduces the sensitivity of the strength of the \ion{N}{ii}
lines to changes in the atmospheric parameters.

For example at (\teff, \logg, \vt) of (23000, 4.0, 0), an equivalent
width of 50 m\AA\ for the \ion{N}{ii} 3995\AA\ line implies a
nitrogen abundance estimate of 7.16~dex. For atmospheric parameters of
(26000, 4.3, 0) and (20000, 3.7, 0), the estimates become 7.19~dex and
7.46~dex respectively, leading to a range of approximately 0.3~dex in
nitrogen abundance estimates for an effective-temperature uncertainty
of $\pm$3\,000\ K.

We have therefore analysed these 13 targets assuming an effective
temperature of 23\,000 K. For a given target, the error in the
effective-temperature estimate will depend on the S/N ratio of the
spectroscopy and the other atmospheric parameters. However, for
effective temperatures higher than 26\,000 K, the \ion{He}{ii} spectra
would normally become observable, whilst for an effective temperature
of less than 20\,000 K, the \ion{Si}{ii} spectra become relatively
strong (for example at \teff=20\,000\,K and \logg=4.0 dex, the \ion{Si}{ii} line at
4131\AA\ has predicted equivalent widths of approximately 35 and
50 m\AA\ for microturbulent velocities of 0 and 5 \kms\ respectively and
an LMC silicon abundance).

This effective-temperature range is consistent with 8 out of 13 of our
targets having a spectral types of B1.5 or B2, for which the LMC
effective-temperature calibration of \citet{tru07} implies effective
temperatures of between 21\,700 and 25\,700\ K for luminosity classes
III to V. Five of our targets have earlier spectral types and we have
therefore searched for evidence of \ion{He}{ii} features in their
observed spectra. For all five stars, no evidence was found for the
line at 4541\AA, whilst for two stars (VFTS\,342 and 719), the line at
4686\AA\ was also not seen. Comparison with theoretical spectra then
implied an effective temperature of less than 26\,000 K for these two
stars.

For two other stars (VFTS\,162 and 501), there was marginal evidence
for a feature at 4686\AA\ and fitting this (with an appropriate
gravity deduced from the hydrogen lines) implied an effective
temperature of approximately 26\,000 K. For the final target,
VFTS\,520, the \ion{He}{ii} feature at 4686\AA\ was more convincing
implying an effective-temperature estimate of 26\,500 K. Hence although
not a formal error estimate, this range of effective temperatures
(20\,000 to 26\,000 K) should be sufficient for most of our targets.

For each target, gravities could then be estimated for this range of
effective temperatures, leading to nitrogen abundance estimates. These
are summarised in Table~\ref{t_Oth} (for an assumed zero
microturbulence) and we note that where:
\begin{enumerate}
\item a range of nitrogen abundance estimates is given, this
  explicitly includes the uncertainty in effective temperature (and
  hence the gravity). The adoption of a larger microturbulence
  (e.g. 5~\kms) would lead to a decrease in these estimates by
  typically less than 0.1~dex.
\item an upper limit for the nitrogen abundance is given, this
  corresponds to the largest estimate found within our range of
  effective temperatures (and corresponding gravities).  The adoption
  of a larger microturbulence (e.g. 5~\kms) would again lead to a
  small decrease in these upper limits.
\end{enumerate}

In summary, the analysis of these targets is less sophisticated as
befits the uncertainties in estimating their effective
temperatures. The nitrogen abundance estimates should therefore be
treated with some caution but they provide a useful supplement to
those given in Table~\ref{t_Atm}. As for the other targets, we would
expect that inclusion of a contribution from the unseen secondary
would again lead to an increase in the estimated nitrogen abundances
of approximately 0.2~dex.

\section{Discussion}\label{s_discussion}

\subsection{Double-lined spectroscopic binaries} \label{s_SB2}

Three of our binaries (VFTS\,240, 520 and 589) were classified by
\citet{eva15} as either SB2? or SB2. VFTS\,240 was excluded from the
present study due to the poor quality of its spectroscopy
(Sect.\ \ref{s_meth}). During our model-atmosphere analysis, evidence was
found for a secondary spectrum in VFTS\,686. We discuss these stars in
more detail below: 

\smallskip\goodbreak 
\noindent
\underline{VFTS\,520:} This was classified as
B1:\,V\,(SB2?) by \citet{eva15} and inspection of the LR02 spectra
showed evidence of a secondary in the \ion{He}{i} lines.  As the
secondary would appear to be fainter this would be consistent with it
having a main-sequence early- to mid-B spectral type.

\smallskip\goodbreak
\noindent
\underline{VFTS\,589:} This was classified as B0.5\,V\,(SB2?)
by \citet{eva15} and in this case evidence for a secondary was
apparent in its LR03 \ion{Si}{iii} spectrum, where all the exposures
were obtained at a single epoch.  By contrast no evidence for a
secondary was found in combined subsets of LR02 exposures obtained at
a given epoch.

For all three \ion{Si}{iii} lines near 4560 \AA, a narrow absorption
feature was found at approximately 1.4 \AA\ to the red of the
absorption line from the primary. Both components have a similar width
($\mbox{FWHM} \simeq 0.87 $\AA\ and 0.89 \AA\ for the secondary and primary,
respectively), implying that the projected rotational velocity of the
secondary was also $\leq$40~\kms. The equivalent widths of the
secondary components were approximately 50\%\ of those of the
primary. Assuming that the equivalents widths in the individual
spectra were the same would then imply a flux ratio of two or a
secondary flux contribution of 33\%.

Similar features are present in the \ion{O}{ii} doublet near
4593 \AA\ in the LR03 spectral region with strengths of approximately
25\%\ of the primary components, implying (making the same assumption
as for the silicon lines) a flux ratio of four or a secondary
contribution of 20\%. Additionally the \ion{He}{i} line at
4713 \AA\ appears to have two components although there is significant
contamination from nebula emission.

For the effective temperature estimated for the primary (see
Table~\ref{t_Atm}), the \ion{Si}{iii} equivalent widths increase with
decreasing effective temperature, whilst those of \ion{O}{ii}
decrease. The relative strengths of the binary components in the
silicon and oxygen spectra would then suggest that the secondary is a
main-sequence star of slightly later type than the primary. In turn
this would then imply a flux contamination from the secondary of
approximately 25\%.

The LR03 spectral range lies close to that of the $B$ photometric
band. Assuming, for example, that the secondary had a spectral type of
B1.5 V, the spectral type versus magnitude calibrations of
\citet{wal72} and \citet{pec13} would imply a $B$-magnitude difference of 0.8-0.9 and a
secondary flux contribution of approximately 30\%, in reasonable
agreement with that estimated above. Finally we note that the upper
limits for the projected rotational velocities of both components
would be consistent with their rotational periods being synchronised
to the orbital period.  

\smallskip\goodbreak
\noindent
\underline{VFTS\,686:} \citet{eva15} classified this target as a
single-lined spectroscopic binary with a B0.7\,III spectral
type. During the model-atmosphere analysis, the presence of a
secondary component was identified in the LR03 combined spectrum. As
for VFTS\,589, no evidence for a secondary was found in combined
subsets of LR02 exposures obtained at different epochs.

The secondary was identified in the LR03 spectrum by broad, shallow
components ($\mbox{FWHM} \simeq 2.8$\AA) that lay approximately 1.7\AA\ to the
red of those of the narrow-lined primary. Gaussian profiles were
fitted to the absorption lines of the \ion{Si}{iii} multiplet near
4560 \AA\ and \ion{O}{ii} doublet near 4593 \AA, together with the
\ion{He}{i} line at 4713 \AA. The equivalents widths of the secondary
were approximately 50\%\ of those of the primary, although there was
considerable scatter, at least in part due to the secondary components
having very small central depth ($\sim$1--2\%).

The use of a Gaussian profile to fit the broad (presumably rotationally
dominated) profiles of the secondary may not be appropriate, Therefore
we have repeated the fitting process assuming rotational broadened
profiles. This led to broadly similar results but also provided estimates of the projected rotational velocity, \vsini, of the secondary that were in the
range 160--210\,\kms.

Because of the difficulty of measuring the spectra of the secondary,
it is not possible to come to any definitive conclusions about its
nature. Possibilities include that it might have a similar spectral
type to the primary but a lower luminosity class, e.g. B0.7\,V or that
it might have a slightly later spectral type. However, the nature of
its spectrum would imply that it is has an early-B spectral type,
whilst it is not possible that the rotational periods of both
components are synchronised to the orbital period.

The identification of these double-lined spectroscopic binaries
provides an insight into our choice of a 20\%\ secondary flux
contribution when preparing Table~\ref{t_Atm}. The results for
VFTS\,589 imply that {\em narrow-lined} secondaries should be
identified (subject to them having a significantly different radial
velocity to that of the primary at the time of observation) if they
contribute more than 20\%\ of the continuum flux. By contrast a
rapidly rotating secondary might not be identified at such flux
levels. Hence it is important that the results presented in
Table~\ref{t_Atm} are considered as representative of the consequences
of an unseen secondary and are not considered as a firm upper limit.

\subsection{Effective temperature and surface gravity} \label{s_TSG}

\begin{figure}
\epsfig{file=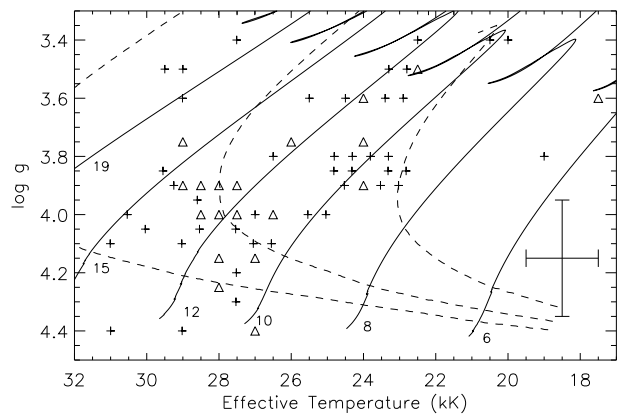,width=9cm, angle=0} 

\caption{Gravity estimates plotted against effective-temperature
  estimates for our binary sample listed in Table~\ref{t_Atm}
  (triangles),  together with representative error bars. 
Also shown is the single-star sample discussed in
  Sect.~\ref{s_N_s_b} (crosses -- some targets have been moved
  slightly in effective temperature or gravity in order to improve
  clarity). Evolutionary models (solid lines) of \citet{bro11a} are
  shown for zero initial rotational velocity together with the initial
  mass (in units of the solar mass). Isochrones (dashed lines) are
  shown for ages of 5, 10 and 20~Myr.}
\label{tempvlogg}
\end{figure}

Figure~\ref{tempvlogg} shows the estimated effective temperatures and
gravities (assuming zero flux contribution from the secondary) for our
binary targets for which a full analysis was undertaken. The targets
discussed in Sect. \ref{s_Other} and summarised in Table~\ref{t_Oth}
have not been included due to the uncertainty in their effective
temperatures. Also shown in Fig.~\ref{tempvlogg} are an equivalent
sample of presumed single targets from the VFTS analysed by
Dufton et al. (in prep.). Both analyses used a similar methodology and hence the
estimates should be comparable. However, it should be noted that
although designated as `single', the sample of Dufton et al. (in prep.) may also
contain some binary systems. As discussed by \citet{dun15} and
\citet{san13}, these will be weighted towards long-period systems
which may have evolved as if they were single.

Also shown are the evolutionary tracks and isochrones of \citet{bro11a} for effectively zero
initial rotational velocity. These were chosen to be consistent with
the observed low projected rotational velocities of our targets. However,
in this part of the HR diagram, the evolutionary tracks and isochrones
are relatively insensitive to the choice of initial rotational
velocity as can be seen from Figs. 5 and 7 of \citet{bro11a}.

Assuming that the primaries of our binary targets have evolved as
single systems, they would appear to have ages of between 5 and
20~Myr. As discussed above, inclusion of a flux contribution from the
secondary would normally increase the estimates of both the effective
temperature and gravity for the primaries. In turn this would normally decrease
the age estimates.  The estimated ages of our targets are consistent with their location outside regions containing the youngest stars in 30 Doradus.

Our coolest target, VFTS\,662, appears to have an age of 40 to 50~Myr greater 
than that of the rest of the sample. The wide-field F775W mosaic of 30~Dor 
taken with the Hubble Space Telescope (HST) in programme
GO-12499 \citep[PI: Lennon; see][]{sab13} shows that VFTS\,662 has a
fainter visual companion \citep{dun15}, with photometry by \citet{sab16}
implying that the companion contributes about 16\%\ of the observed flux
in our spectra. Additionally this target could have experienced
interaction with its spectroscopic secondary making a comparison with single 
star models invalid.

The sample of presumed single stars lies in a similar part of
Fig.~\ref{tempvlogg} to that occupied by the binaries. In turn this
leads to a similar range for their ages and masses, although the
single-star sample may contain more higher-mass objects. This sample
contains one relatively cool star, VFTS\,273, which also appears to
have an anomalously large evolutionary age estimate; further
discussion of this object will be deferred to Dufton et al. (in prep.).

In summary the single- and binary-star samples cover a similar range
of atmospheric parameters making them suitable for the comparison of
their nitrogen abundances as discussed in Sect.~\ref{s_N_d}.

\subsection{HR diagram} \label{s_HR}

\begin{figure}
\epsfig{file=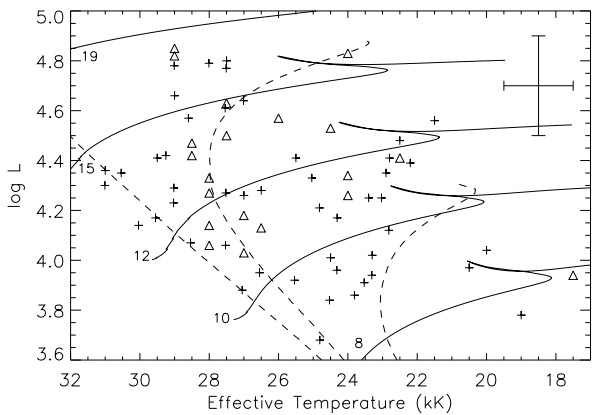,width=9cm, angle=0} 

\caption{Luminosity estimates (in units of the solar luminosity)
  plotted against effective-temperature estimates for our binary
  sample in Table~\ref{t_Atm} (triangles), together with representative error bars. 
  Also shown are the
  single-star sample discussed in Sect.~\ref{s_N_s_b} (crosses --
  some targets have been moved slightly in effective temperature or
  gravity in order to improve clarity). Evolutionary models (solid
  lines) of \citet{bro11a} are shown for zero initial rotational
  velocity together with the initial mass (in units of the solar
  mass). Isochrones (dashed lines) are shown for ages of 5, 10 and 20~Myr.}
\label{tempvlum}
\end{figure}

Luminosities have been estimated for the binary systems for which a
full analysis was undertaken together with those for the equivalent
sample of presumed single targets from the VFTS discussed in
Sect.~\ref{s_TSG}. Interstellar extinctions have been estimated from
observed $B-V$ colours provided in Paper~I, together with the
intrinsic-colour--spectral-type calibration of \citet{weg94} plus
$R_{V} = 3.5$ from \citet{dor13}. Bolometric corrections were obtained
from LMC-metallicity {\sc tlusty} models \citep{lan07} for the primary
component, and together with an adopted distance modulus of 18{\fm}5
for the LMC permitted the systemic luminosity to be estimated.

For one target, VFTS\,017, optical photometry was not available and we
used the HST photometry of \citet{sab16} in the F555W band, together
with a E$(B-V)=0.31$ (average of that found for other B-type stars in
VFTS) to estimate a logarithmic luminosity of 4.85 (in units of the
solar luminosity). We note that this agrees well with the estimate for
VFTS\,534, which has similar atmospheric parameters. The estimates are
listed in Table~\ref{t_Targets}.

The values for the binary sample should be treated with some
caution. For example, a significant secondary contribution of the flux
in the $V$\ photometric band would lead to an overestimation of the {\em
  primary's} luminosity. For a 20\%\ secondary flux contribution, this
would translate to an overestimate of approximately
0.1~dex. Additionally the presence of a cooler secondary could lead to
an overestimation of the reddening leading in turn to an
overestimation of the luminosity. However, this effect is likely to be
relatively small as any secondary making a significant contribution to
the total flux would have a similar colour to the primary. Given the uncertainty in the contribution of the secondary to the total luminosity, a conservative error estimate of $\pm0.2$ dex has been adopted.

The HR diagram for both the binary and single samples is shown in
Fig.~\ref{tempvlum}, together with the same evolutionary tracks and
isochrones of \citet{bro11a} as shown in Fig.~\ref{tempvlogg}. These
lead to age estimates similar to those found in Sect.~\ref{s_TSG},
with for example, the binaries having a typical age of 10 Myr. One
target, VFTS\,662, again has an age of more than 20 Myr. The single stars 
cover a similar range of age estimates. In this case, three targets appear to 
have age estimates greater than 20 Myr, which would be consistent with the larger sample size.

\subsection{Nitrogen abundance} \label{s_N_d}

\subsubsection{General characteristics} \label{GC}

\begin{figure}
\epsfig{file=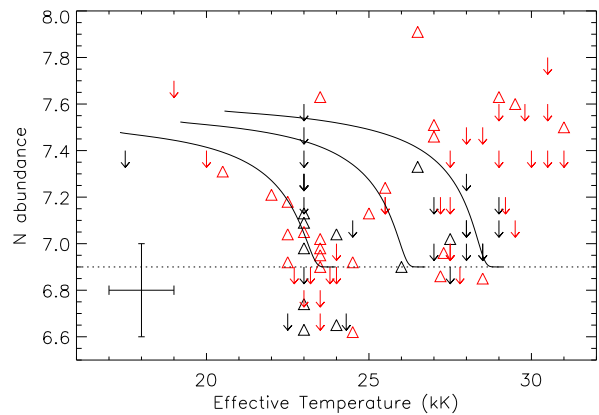,width=9cm, angle=0}

\caption{ Nitrogen abundance estimates (triangles) and upper limits (arrows) assuming no secondary flux contribution are plotted against effective-temperature estimates for our  binary sample (black symbols).  The representative error bars are for targets listed in Table \ref{t_Atm}, with those for targets from Table \ref{t_Oth} being larger - see text for details. 
Also shown are the single-star   sample discussed in Sect.~\ref{s_N_s_b} (red symbols -- some targets have been moved slightly in effective temperature in order to improve clarity). The  dotted line represents a baseline LMC abundance of  6.9~dex. The solid lines are evolutionary models from \citet{bro11a} with initial masses of 8, 10 and 12\Msun\ 
and an initial equatorial rotational velocity of approximately 230 \kms.}
\label{N_teff}
\end{figure} 

The LMC baseline nitrogen abundance has been estimated from
observations of both \ion{H}{ii} regions \citep[see, for
  example,][]{kur98, gar99} and early-type stars \citep[see, for
  example,][]{kor02, kor05, hun07, tru07}. The different studies are
in good agreement and imply a value of approximately 6.9~dex, which
will be adopted here.

Our nitrogen abundances estimates for the fully analysed binary sample
(see Table~\ref{t_Atm}) are in general similar to this baseline LMC
abundance. For example, assuming no flux contribution from the
secondary, four out of the five targets with specific estimates show
an enhancement of less than 0.2~dex. For those targets with upper
limits, eleven out of fifteen have enhancements of 0.2~dex or less
(with, of course, it being possible that the other four targets also
have small enhancements). The results for the targets where no  reliable
effective temperature could be estimated (see Table~\ref{t_Oth}) are
also compatible with relatively modest enhancements. For example, for
the five stars where a range in nitrogen abundances could be
estimated, three show effectively no enhancements, whilst the other
two show enhancements of approximately 0.1--0.3~dex. For both samples,
including a contribution from a secondary would increase the actual or
possible enhancements but they would still remain relatively modest.

As discussed in Sect.\ \ref{s_meth}, our grid of model atmospheres assumed a normal helium and hydrogen abundance ratio. The LMC evolutionary models of \citet{bro11a} appropriate to B-type stars indicate that even for a nitrogen enhancement of 1.0 dex (larger than that observed in our sample -- see Tables \ref{t_Atm} and \ref{t_Oth}), the change in the helium abundance is typically only  0.03 dex. Hence our assumption of a normal helium abundance is unlikely to be a significant source of error.

Both the Tarantula and FLAMES-I samples were restricted to stars with
low {\em projected} rotational velocities. We would expect the
majority of these stars to also have small {\em equatorial} rotational
velocities for the following reasons:
\begin{enumerate}
\item Assuming a random distribution of axes of inclination, the
  probability of observing small angles of inclination (and hence
  small $\sin i$) is low. For example a $\sin i\leq 0.25$\ would only
  occur in approximately 3\%\ of targets.
\item The identification of binaries will be biased towards systems
  with large orbital angles of inclination as this will lead to a
  large range in radial-velocity variations \citep{dun15}. Surmising that the axes of the orbital and rotational motions are aligned would again favour larger rotational angles of inclination.
\end{enumerate}

The relatively small enhancements in the nitrogen abundances found in the two binary samples would then be consistent with the primaries having
evolved as slowly rotating single stars, which have experienced little
rotational mixing between the stellar core and envelope. Indeed as
discussed by \citet{deM09, deM11a}, pre-interaction binaries `may provide the most stringent test cases for single stellar models'. 

For the Tarantula sample, the range of radial velocity variations for each primary, \dvr, is listed in Table \ref{tab.basics}. These cover a significant range from approximately 20 to 180 \kms. Six of the targets have estimates of \dvr\ of over 100 \kms and might be expected to be amongst the most tightly bound systems. Inspection of Tables \ref{t_Atm} and \ref{t_Oth} indicate that (assuming the secondary contributes zero flux) four targets (VFTS\,342, 501, 589, 888) have nitrogen enhancements of 0.1 dex or less; two (VFTS\,299, 520) have upper limits on the nitrogen abundance of $\leq$7.3 dex, implying a {\em maximum} enhancement of 0.4 dex. Hence for these presumably closely bound systems, there is no evdence of substantial nitrogen abundance enhancements.   This provides constraints on the combined effect of the physical processes that would lead to such enhancements.  For example, it appears that these  closely bound systems binaries have not yet interacted through mass transfer, consistent with the predictions by \citet{deM11a, deM14}. Additionally it constrains the combined effect of any further mixing processes that may have operated. Specifically, it implies that the effects of rotationally induced mixing have been limited  \citep{deM09,bro11b}.

We can compare our nitrogen abundances with those found for the binary VFTS supergiants analysed by \citet{mce14}.  Assuming that single star evolutionary models are appropriate, these supergiants will normally have evolved from O-type main sequence stars and hence will {\em not} be the descendants of our current sample. \citet{mce14} found some tentative evidence that low (i.e. near baseline) nitrogen abundances were more prevalent in their B-type primaries. These objects may have evolved from O-type main sequence stars that are analogous to the low nitrogen abundance primaries found in our sample. Additionally \citet{mce14} identified a small number of binary supergiants with very high nitrogen abundances ($\sim$8.0 dex) that showed evidence for being post-interaction systems.

\subsubsection{Comparison with single stars} \label{s_N_s_b}

We plot the nitrogen abundance estimates for our binary sample as a
function of effective temperature in Fig.~\ref{N_teff}. The estimates
assuming negligible flux contribution from the secondary were adopted;
adopting a 20\%\ secondary contribution would increase the estimates
by typically 0.1--0.2~dex (see Table~\ref{t_Atm}). Also shown are LMC evolutionary tracks for stars with initial masses of
8, 10 and 12\Msun\ and an initial equatorial velocity of approximately 230 \kms taken from \citet{bro11a}. These tracks stretch from the zero age main sequence to when the model has a surface gravity, \logg$\sim$ 3.4 dex, consistent with the gravity range of our sample. As discussed in
Sect.~\ref{s_TSG}, an analysis of an equivalent sample of apparently single stars
in the VFTS has been carried out by
Dufton et al. (in prep.) using similar methods to those used here. These are also plotted in Fig.~\ref{N_teff} and were selected using the same criteria  as for the binary sample, i.e. \vsini$\leq$ 80\,\kms and excluding supergiants discussed by \citet{mce14}.

For the binary sample, there are a significant number of stars plotted
with $\teff=23\,000$\,K. These correspond to the targets listed in
Table~\ref{t_Oth} and in reality will occupy the effective-temperature
range ${\sim}$20\,000--26\,000\,K. Additionally, many of the abundance
estimates are upper limits, implying that Fig.~\ref{N_teff}
should be interpreted carefully. 

Nine targets (5 binaries, 4 `single' stars) appear to have nitrogen
abundances that are 0.1--0.3~dex below the adopted baseline
abundance;  this may simply be due to random uncertainties, estimated in
Sect.~\ref{s_N} as being of the order of 0.2--0.3~dex. Additionally,
Table~\ref{t_Atm} implies that inclusion of a secondary flux would
increase our estimates by 0.1--0.2~dex. Indeed the same effect could
be affecting our single-star estimates if there were undiscovered
binaries. Hence we do not believe that these stars stars provide any
convincing evidence for nitrogen abundances that are truly below our
adopted baseline.  

Both samples appear to contain significant numbers of stars that have
nitrogen abundance estimates close to the assumed baseline abundance
of the LMC. Additionally, targets with modest nitrogen enhancements
($\le 0.5$~dex) are also observed. As discussed in Sect \ref{GC}, this
would be compatible with their evolution as slowly rotating stars with
little rotational mixing.
One possible difference between the samples is that the single-star
sample may contain targets with larger nitrogen abundance enhancements
(of up to 1.1~dex). For example, four single stars have estimated abundances
that are greater than both the detections and even the highest upper
limit found in the binary sample. Such targets have also been found in
other presumably single-star samples in the Magellanic Cloud
\citep{hun07, hun08b, hun09a, tru07}.

The single-star sample consists of 54 targets of which six have nitrogen enhancements of more than 0.6 dex. As discussed in Sect. \ref{s_TSG}, it contains more high mass and hence high effective temperature targets. In particular it has 9 targets that have effective temperatures, \teff$>$29\,000\,K, which is the maximum temperature of the binary sample. Two of these targets show nitrogen enhancements of more than 0.6 dex, leaving 4  (out of 45) targets with significantly enhanced nitrogen in the effective temperature range covered by the binaries. Hence the lack of such objects in the binary sample may at least in part be due the different effective temperature ranges that have been sampled.

To examine more rigorously the possible differences between binary and single-star
samples requires statistical tools.  Standard tests
for comparing two univariate samples, such as the Kolmogorov-Smirnov
and Kuiper tests, are ill-suited to our nitrogen abundance datasets because
of the large numbers of upper limits (15/20 and 30/54 for the binary
and single-star samples, respectively).  However, alternative tools
exist for treating data that are randomly censored.\footnote{
`Randomly censored' implies that the probability of the measurement
of a given target being an upper limit is independent of the actual value. This may not always be the case here as we would expect to preferentially obtain estimates for targets with large nitrogen abundances.
Many of the techniques for characterizing such data were originally developed principally in the field of
`survival statistics', and were first introduced to the astrophysics
community by \citet{fei85} and by \citet{sch85}.}
To the extent that detections and limits are intermingled, our
abundance results can be approximated  as randomly censored datasets (as opposed
to being truncated, whereby all detections would lie above a
lower-limit cutoff).

We have used the \textsc{asurv} package (v1.2; \citealt{lav92}), which
implements several tests for censored data, to investigate
whether the null hypothesis of no differences between the binary and
single-star samples can be rejected with appropriate levels of
confidence.  Results are given in Table~\ref{t_idh} for both secondary flux contributions of 0\% and 20\%.   The two-sample
(Wilcoxon \& log\-rank) tests are nonparametric comparisons of the
distributions, and are moderately insensitive to the details of the
censoring (see, e.g., \citealt{fei85}).  None provides any strong
evidence for differences between results for the binary and
single-star samples, particularly if a modest secondary-flux
contribution is accounted for.   

The Kaplan-Meier estimator, used to estimate the means of the
distributions, is more sensitively dependent on the assumption of truly
random censoring.  Given this dependence, our interpretation of the
results in Table~\ref{t_idh} is again that there is no compelling
evidence for any overall differences in the single-star and binary nitrogen
abundances.

In Sect. \ref{s_N}, the  \ion{He}{i} line at 4009\AA\  was found to be consistent with secondary flux contribution of between 0 and 20\% for the binary sample. Hence we would expect that the use of actual nitrogen abundances (rather than our representative estimates) for the primaries in our binary sample would have led to similar results to those found in Table \ref{t_idh}.

As discussed above, the single star sample contains stars with high effective temperatures ($>$29\,000\,K) that are not present in the binary sample. We have therefore repeated these tests for those targets within the range, 26\,000$\leq$\teff$\leq$29\,000\,K; the choice of a limited range of effective temperatures should also lead to the two samples having similar ages and masses. For the binary sample, this lead to 14 targets (11 having upper limits for the nitrogen abundance ) for a zero secondary flux contribution and 11 targets (and 7 upper limits) for a 20\% contribution. For the single star sample, there were 18 targets with 11 upper limits.

The statistical tests are again summarized in Table \ref{t_idh} and in general show no evidence of any difference between the two samples. Indeed the probabilities that the two parent distributions are indistinguishable have increased at least in part due to the smaller sample sizes. Additionally the estimated mean nitrogen abundances of the two samples again show no significant differences.

Therefore we conclude that the statistical tests are consistent with the null 
hypothesis (that there are no differences between the binary and
single-star samples), although in some cases the probabilities listed in Table \ref{t_idh} are relatively small.
Indeed Fig.\ \ref{N_teff} does imply an absence of targets with high nitrogen abundances in the binary
sample. As discussed in Sect.~\ref{GC}, this would be consistent with them having evolved as effectively single stars with low rotational velocities.

The relatively large nitrogen abundances in some of the single stars could then arise from rapidly rotating stars viewed at low angles of incidence. However such viewing angles are unlikely (for example, a value of $\sin i \le 0.1$ will only occur in 0.5\%\ of targets with randomly aligned rotation axes).  This has lead other authors \citep{hun08b, bro11b, gri16} to question whether all the low-\vsini\ single stars with high nitrogen abundances found in other single-star samples can be rapid rotators viewed at low angles of incidence. Indeed other mechanisms \citep[for example, binary mergers, see][]{deM14} may also be required and this will be considered in detail in Dufton et al. (in prep.).

The binaries discussed here, together with those for the VFTS binary supergiants discussed by \citet{mce14} are the first  significant binary samples to be analysed using model atmosphere techniques. As such they should be considered as pathfinder analyses especially given the difficulty, for example, in allowing for the spectral contributions of unseen secondaries. They should provide a motivation for spectroscopic abundance analysis of larger samples of binary stars, especially given binaries dominate the massive star population \citep{san12}.

\begin{table*}

\caption{Statistical comparisons of the binary and single-star
  samples.  The first four entries (Wilcoxon and log\-rank tests) give
  the probabilities that the censored single-star and binary
  distributions of nitrogen abundances are indistinguishable, for two
  assumed values of the secondary's flux contribution to the binary-star
  spectra.   The last two entries give the mean abundances evaluated
  using the Kaplan-Meier estimator (see Sect.\ \ref{s_N_s_b} for details).  The 'Full' sample is taken from Table \ref{t_Atm}, whilst the 'Limited'  sample is for the effective temperature range 26\,000-29\,000\,K.}
\label{t_idh}

\begin{center}
\begin{tabular}{lcccc}
\hline\hline
$\qquad$Test & \multicolumn{2}{c}{Full} & \multicolumn{2}{c}{Limited \teff\ range} \\
     & 0\%\ & 20\%\ \\
\hline
Gehan generalized Wilcoxon & 0.06 & 0.34 & 0.22 & 0.79\\
Peto \& Peto Wilcoxon      & 0.06 & 0.40 & 0.43  & 0.77\\
Peto \& Prentice Wilcoxon  & 0.06 & 0.36 & 0.41 & 0.80\\
log\-rank                    & 0.11 & 0.44 & 0.57 & 0.54\\
\\
\multicolumn{3}{l}{Mean N abundances (Kaplan-Meier estimator):}\\
Binary sample              & $6.79 \pm 0.06$ & $6.92 \pm 0.07$ & $6.95\pm0.03$  & $7.12\pm0.05$ \\
Single-star sample         &\multicolumn{2}{c}{$6.98 \pm 0.05$} & \multicolumn{2}{c}{$7.06 \pm 0.08$}\\
\hline
\end{tabular}
\end{center}
\end{table*}

\begin{acknowledgements}
Based on observations at the European Southern Observatory Very Large Telescope in programme 182.D-0222. SdM acknowledges support by a Marie Sklodowska-Curie Action (H2020 MSCA-IF-2014, project BinCosmos, id 661502). RG would like to thank the Institute of Physics and the Nuffield Foundation for helping fund this work through their Undergraduate Research Bursary program.  
\end{acknowledgements}

\bibliography{VFTS_low_vsini_binary_ref.bib}

\newcommand{\noop}[1]{}
\begin{thebibliography}{77}
\expandafter\ifx\csname natexlab\endcsname\relax\def\natexlab#1{#1}\fi

\bibitem[{{Aerts} {et~al.}(2014){Aerts}, {Molenberghs}, {Kenward}, \&
  {Neiner}}]{aer14a}
{Aerts}, C., {Molenberghs}, G., {Kenward}, M.~G., \& {Neiner}, C. 2014, \apj,
  781, 88

\bibitem[{{Brott} {et~al.}(2011{\natexlab{a}}){Brott}, {de Mink}, {Cantiello},
  {Langer}, {de Koter}, {Evans}, {Hunter}, {Trundle}, \& {Vink}}]{bro11a}
{Brott}, I., {de Mink}, S.~E., {Cantiello}, M., {et~al.} 2011{\natexlab{a}},
  \aap, 530, A115

\bibitem[{{Brott} {et~al.}(2011{\natexlab{b}}){Brott}, {Evans}, {Hunter}, {de
  Koter}, {Langer}, {Dufton}, {Cantiello}, {Trundle}, {Lennon}, {de Mink},
  {Yoon}, \& {Anders}}]{bro11b}
{Brott}, I., {Evans}, C.~J., {Hunter}, I., {et~al.} 2011{\natexlab{b}}, \aap,
  530, A116

\bibitem[{{Chiosi} \& {Maeder}(1986)}]{chi86}
{Chiosi}, C. \& {Maeder}, A. 1986, \araa, 24, 329

\bibitem[{{de Mink} {et~al.}(2009){de Mink}, {Cantiello}, {Langer}, {Pols},
  {Brott}, \& {Yoon}}]{deM09}
{de Mink}, S.~E., {Cantiello}, M., {Langer}, N., {et~al.} 2009, \aap, 497, 243

\bibitem[{{de Mink} {et~al.}(2011){de Mink}, {Langer}, \& {Izzard}}]{deM11a}
{de Mink}, S.~E., {Langer}, N., \& {Izzard}, R.~G. 2011, Bulletin de la Societe
  Royale des Sciences de Liege, 80, 543

\bibitem[{{de Mink} {et~al.}(2014){de Mink}, {Sana}, {Langer}, {Izzard}, \&
  {Schneider}}]{deM14}
{de Mink}, S.~E., {Sana}, H., {Langer}, N., {Izzard}, R.~G., \& {Schneider},
  F.~R.~N. 2014, \apj, 782, 7

\bibitem[{{Donati} \& {Landstreet}(2009)}]{don09}
{Donati}, J.-F. \& {Landstreet}, J.~D. 2009, \araa, 47, 333

\bibitem[{{Doran} {et~al.}(2013){Doran}, {Crowther}, {de Koter}, {Evans},
  {McEvoy}, {Walborn}, {Bastian}, {Bestenlehner}, {Gr{\"a}fener}, {Herrero},
  {K{\"o}hler}, {Ma{\'{\i}}z Apell{\'a}niz}, {Najarro}, {Puls}, {Sana},
  {Schneider}, {Taylor}, {van Loon}, \& {Vink}}]{dor13}
{Doran}, E.~I., {Crowther}, P.~A., {de Koter}, A., {et~al.} 2013, \aap, 558,
  A134

\bibitem[{{Dufton} {et~al.}(2013){Dufton}, {Langer}, {Dunstall}, {Evans},
  {Brott}, {de Mink}, {Howarth}, {Kennedy}, {McEvoy}, {Potter},
  {Ram{\'{\i}}rez-Agudelo}, {Sana}, {Sim{\'o}n-D{\'{\i}}az}, {Taylor}, \&
  {Vink}}]{duf12}
{Dufton}, P.~L., {Langer}, N., {Dunstall}, P.~R., {et~al.} 2013, \aap, 550,
  A109

\bibitem[{{Dufton} {et~al.}(2006){Dufton}, {Ryans}, {Sim{\'o}n-D{\'{\i}}az},
  {Trundle}, \& {Lennon}}]{dufsmc06}
{Dufton}, P.~L., {Ryans}, R.~S.~I., {Sim{\'o}n-D{\'{\i}}az}, S., {Trundle}, C.,
  \& {Lennon}, D.~J. 2006, \aap, 451, 603

\bibitem[{{Dufton} {et~al.}(2005){Dufton}, {Ryans}, {Trundle}, {Lennon},
  {Hubeny}, {Lanz}, \& {Allende Prieto}}]{duf05}
{Dufton}, P.~L., {Ryans}, R.~S.~I., {Trundle}, C., {et~al.} 2005, \aap, 434,
  1125

\bibitem[{{Dunstall} {et~al.}(2011){Dunstall}, {Brott}, {Dufton}, {Lennon},
  {Evans}, {Smartt}, \& {Hunter}}]{dun11}
{Dunstall}, P.~R., {Brott}, I., {Dufton}, P.~L., {et~al.} 2011, \aap, 536, A65

\bibitem[{{Dunstall} {et~al.}(2015){Dunstall}, {Dufton}, {Sana}, {Evans},
  {Howarth}, {Sim{\'o}n-D{\'{\i}}az}, {de Mink}, {Langer}, {Ma{\'{\i}}z
  Apell{\'a}niz}, \& {Taylor}}]{dun15}
{Dunstall}, P.~R., {Dufton}, P.~L., {Sana}, H., {et~al.} 2015, \aap, 580, A93

\bibitem[{{Ekstr{\"o}m} {et~al.}(2012){Ekstr{\"o}m}, {Georgy}, {Eggenberger},
  {Meynet}, {Mowlavi}, {Wyttenbach}, {Granada}, {Decressin}, {Hirschi},
  {Frischknecht}, {Charbonnel}, \& {Maeder}}]{eka12}
{Ekstr{\"o}m}, S., {Georgy}, C., {Eggenberger}, P., {et~al.} 2012, \aap, 537,
  A146

\bibitem[{{Evans} {et~al.}(2008){Evans}, {Hunter}, {Smartt}, {Lennon}, {de
  Koter}, {Mokiem}, {Trundle}, {Dufton}, {Ryans}, {Puls}, {Vink}, {Herrero},
  {Sim{\'o}n-D{\'{\i}}az}, {Langer}, \& {Brott}}]{eva08}
{Evans}, C., {Hunter}, I., {Smartt}, S., {et~al.} 2008, The Messenger, 131, 25

\bibitem[{{Evans} {et~al.}(2011{\natexlab{a}}){Evans}, {Davies}, {Kudritzki},
  {Puech}, {Yang}, {Cuby}, {Figer}, {Lehnert}, {Morris}, \& {Rousset}}]{eva11a}
{Evans}, C.~J., {Davies}, B., {Kudritzki}, R.-P., {et~al.} 2011{\natexlab{a}},
  \aap, 527, A50

\bibitem[{{Evans} {et~al.}(2015){Evans}, {Kennedy}, {Dufton}, {Howarth},
  {Walborn}, {Markova}, {Clark}, {de Mink}, {de Koter}, {Dunstall},
  {H{\'e}nault-Brunet}, {Ma{\'{\i}}z Apell{\'a}niz}, {McEvoy}, {Sana},
  {Sim{\'o}n-D{\'{\i}}az}, {Taylor}, \& {Vink}}]{eva15}
{Evans}, C.~J., {Kennedy}, M.~B., {Dufton}, P.~L., {et~al.} 2015, \aap, 574,
  A13

\bibitem[{{Evans} {et~al.}(2011{\natexlab{b}}){Evans}, {Taylor},
  {H{\'e}nault-Brunet}, {Sana}, {de Koter}, {Sim{\'o}n-D{\'{\i}}az}, {Carraro},
  {Bagnoli}, {Bastian}, {Bestenlehner}, {Bonanos}, {Bressert}, {Brott},
  {Campbell}, {Cantiello}, {Clark}, {Costa}, {Crowther}, {de Mink}, {Doran},
  {Dufton}, {Dunstall}, {Friedrich}, {Garcia}, {Gieles}, {Gr{\"a}fener},
  {Herrero}, {Howarth}, {Izzard}, {Langer}, {Lennon}, {Ma{\'{\i}}z
  Apell{\'a}niz}, {Markova}, {Najarro}, {Puls}, {Ramirez},
  {Sab{\'{\i}}n-Sanjuli{\'a}n}, {Smartt}, {Stroud}, {van Loon}, {Vink}, \&
  {Walborn}}]{eva11}
{Evans}, C.~J., {Taylor}, W.~D., {H{\'e}nault-Brunet}, V., {et~al.}
  2011{\natexlab{b}}, \aap, 530, A108

\bibitem[{{Feigelson} \& {Nelson}(1985)}]{fei85}
{Feigelson}, E.~D. \& {Nelson}, P.~I. 1985, \apj, 293, 192

\bibitem[{{Fraser} {et~al.}(2010){Fraser}, {Dufton}, {Hunter}, \&
  {Ryans}}]{fra10}
{Fraser}, M., {Dufton}, P.~L., {Hunter}, I., \& {Ryans}, R.~S.~I. 2010, \mnras,
  404, 1306

\bibitem[{{Frischknecht} {et~al.}(2010){Frischknecht}, {Hirschi}, {Meynet},
  {Ekstr{\"o}m}, {Georgy}, {Rauscher}, {Winteler}, \& {Thielemann}}]{fri10}
{Frischknecht}, U., {Hirschi}, R., {Meynet}, G., {et~al.} 2010, \aap, 522, A39

\bibitem[{{Garnett}(1999)}]{gar99}
{Garnett}, D.~R. 1999, in IAU Symposium, Vol. 190, New Views of the Magellanic
  Clouds, ed. {Y.-H.~Chu, N.~Suntzeff, J.~Hesser, \& D.~Bohlender}, 266

\bibitem[{{Georgy} {et~al.}(2013){Georgy}, {Ekstr{\"o}m}, {Eggenberger},
  {Meynet}, {Haemmerl{\'e}}, {Maeder}, {Granada}, {Groh}, {Hirschi}, {Mowlavi},
  {Yusof}, {Charbonnel}, {Decressin}, \& {Barblan}}]{geo13}
{Georgy}, C., {Ekstr{\"o}m}, S., {Eggenberger}, P., {et~al.} 2013, \aap, 558,
  A103

\bibitem[{{Gonz{\'a}lez} \& {Levato}(2006)}]{gon06}
{Gonz{\'a}lez}, J.~F. \& {Levato}, H. 2006, \aap, 448, 283

\bibitem[{{Gray}(2005)}]{gra05}
{Gray}, D.~F. 2005, {The Observation and Analysis of Stellar Photospheres}

\bibitem[{{Grin} {et~al.}(2016){Grin}, {Ramirez-Agudelo}, {de Koter}, {Sana},
  {Puls}, {Brott}, {Crowther}, {Dufton}, {Evans}, {Graefener}, {Herrero},
  {Langer}, {Lennon}, {van Loon}, {Markova}, {de Mink}, {Najarro}, {Schneider},
  {Taylor}, {Tramper}, {Vink}, \& {Walborn}}]{gri16}
{Grin}, N.~J., {Ramirez-Agudelo}, O.~H., {de Koter}, A., {et~al.} 2016, ArXiv
  e-prints

\bibitem[{{Heger} \& {Langer}(2000)}]{heg00b}
{Heger}, A. \& {Langer}, N. 2000, \apj, 544, 1016

\bibitem[{{Heger} {et~al.}(2000){Heger}, {Langer}, \& {Woosley}}]{heg00a}
{Heger}, A., {Langer}, N., \& {Woosley}, S.~E. 2000, \apj, 528, 368

\bibitem[{{Hirschi} {et~al.}(2004){Hirschi}, {Meynet}, \& {Maeder}}]{hir04}
{Hirschi}, R., {Meynet}, G., \& {Maeder}, A. 2004, \aap, 425, 649

\bibitem[{{Howarth} {et~al.}(2015){Howarth}, {Dufton}, {Dunstall}, {Evans},
  {Almeida}, {Bonanos}, {Clark}, {Langer}, {Sana}, {Sim{\'o}n-D{\'{\i}}az},
  {Soszy{\'n}ski}, \& {Taylor}}]{how15}
{Howarth}, I.~D., {Dufton}, P.~L., {Dunstall}, P.~R., {et~al.} 2015, \aap, 582,
  A73

\bibitem[{{Hubeny}(1988)}]{hub88}
{Hubeny}, I. 1988, Computer Physics Communications, 52, 103

\bibitem[{{Hubeny} {et~al.}(1998){Hubeny}, {Heap}, \& {Lanz}}]{hub98}
{Hubeny}, I., {Heap}, S.~R., \& {Lanz}, T. 1998, in Astronomical Society of the
  Pacific Conference Series, Vol. 131, Properties of Hot Luminous Stars, ed.
  {I.~Howarth}, 108--+

\bibitem[{{Hubeny} \& {Lanz}(1995)}]{hub95}
{Hubeny}, I. \& {Lanz}, T. 1995, \apj, 439, 875

\bibitem[{{Hunter} {et~al.}(2009){Hunter}, {Brott}, {Langer}, {Lennon},
  {Dufton}, {Howarth}, {Ryans}, {Trundle}, {Evans}, {de Koter}, \&
  {Smartt}}]{hun09a}
{Hunter}, I., {Brott}, I., {Langer}, N., {et~al.} 2009, \aap, 496, 841

\bibitem[{{Hunter} {et~al.}(2008){Hunter}, {Brott}, {Lennon}, {Langer},
  {Dufton}, {Trundle}, {Smartt}, {de Koter}, {Evans}, \& {Ryans}}]{hun08b}
{Hunter}, I., {Brott}, I., {Lennon}, D.~J., {et~al.} 2008, \apjl, 676, L29

\bibitem[{{Hunter} {et~al.}(2007){Hunter}, {Dufton}, {Smartt}, {Ryans},
  {Evans}, {Lennon}, {Trundle}, {Hubeny}, \& {Lanz}}]{hun07}
{Hunter}, I., {Dufton}, P.~L., {Smartt}, S.~J., {et~al.} 2007, \aap, 466, 277

\bibitem[{{Kobulnicky} {et~al.}(2014){Kobulnicky}, {Kiminki}, {Lundquist},
  {Burke}, {Chapman}, {Keller}, {Lester}, {Rolen}, {Topel}, {Bhattacharjee},
  {Smullen}, {Vargas {\'A}lvarez}, {Runnoe}, {Dale}, \& {Brotherton}}]{kob14}
{Kobulnicky}, H.~A., {Kiminki}, D.~C., {Lundquist}, M.~J., {et~al.} 2014,
  \apjs, 213, 34

\bibitem[{{Korn} {et~al.}(2002){Korn}, {Keller}, {Kaufer}, {Langer},
  {Przybilla}, {Stahl}, \& {Wolf}}]{kor02}
{Korn}, A.~J., {Keller}, S.~C., {Kaufer}, A., {et~al.} 2002, \aap, 385, 143

\bibitem[{{Korn} {et~al.}(2005){Korn}, {Nieva}, {Daflon}, \& {Cunha}}]{kor05}
{Korn}, A.~J., {Nieva}, M.~F., {Daflon}, S., \& {Cunha}, K. 2005, \apj, 633,
  899

\bibitem[{{Kudritzki} {et~al.}(2016){Kudritzki}, {Castro}, {Urbaneja}, {Ho},
  {Bresolin}, {Gieren}, {Pietrzy{\'n}ski}, \& {Przybilla}}]{kur16}
{Kudritzki}, R.~P., {Castro}, N., {Urbaneja}, M.~A., {et~al.} 2016, \apj, 829,
  70

\bibitem[{{Kurt} \& {Dufour}(1998)}]{kur98}
{Kurt}, C.~M. \& {Dufour}, R.~J. 1998, in Revista Mexicana de Astronomia y
  Astrofisica Conference Series, Vol.~7, Revista Mexicana de Astronomia y
  Astrofisica Conference Series, ed. R.~J. {Dufour} \& S.~{Torres-Peimbert},
  202

\bibitem[{{Langer}(2012)}]{lan12}
{Langer}, N. 2012, ARAA, 50, 107

\bibitem[{{Lanz} \& {Hubeny}(2007)}]{lan07}
{Lanz}, T. \& {Hubeny}, I. 2007, \apjs, 169, 83

\bibitem[{{LaValley} {et~al.}(1992){LaValley}, {Isobe}, \& {Feigelson}}]{lav92}
{LaValley}, M., {Isobe}, T., \& {Feigelson}, E. 1992, in Astronomical Society
  of the Pacific Conference Series, Vol.~25, Astronomical Data Analysis
  Software and Systems I, ed. D.~M. {Worrall}, C.~{Biemesderfer}, \&
  J.~{Barnes}, 245

\bibitem[{{Lefever} {et~al.}(2007){Lefever}, {Puls}, \& {Aerts}}]{lef07}
{Lefever}, K., {Puls}, J., \& {Aerts}, C. 2007, \aap, 463, 1093

\bibitem[{{Maeder}(1987)}]{mae87}
{Maeder}, A. 1987, \aap, 173, 247

\bibitem[{{Maeder}(2009)}]{mae09}
{Maeder}, A. 2009, {Physics, Formation and Evolution of Rotating Stars}

\bibitem[{{Mahy} {et~al.}(2009){Mahy}, {Naz{\'e}}, {Rauw}, {Gosset}, {De
  Becker}, {Sana}, \& {Eenens}}]{mah09}
{Mahy}, L., {Naz{\'e}}, Y., {Rauw}, G., {et~al.} 2009, \aap, 502, 937

\bibitem[{{Mahy} {et~al.}(2013){Mahy}, {Rauw}, {De Becker}, {Eenens}, \&
  {Flores}}]{mah13}
{Mahy}, L., {Rauw}, G., {De Becker}, M., {Eenens}, P., \& {Flores}, C.~A. 2013,
  \aap, 550, A27

\bibitem[{{Markova} \& {Puls}(2008)}]{mar07}
{Markova}, N. \& {Puls}, J. 2008, \aap, 478, 823

\bibitem[{{Mason} {et~al.}(2009){Mason}, {Hartkopf}, {Gies}, {Henry}, \&
  {Helsel}}]{mas09}
{Mason}, B.~D., {Hartkopf}, W.~I., {Gies}, D.~R., {Henry}, T.~J., \& {Helsel},
  J.~W. 2009, \aj, 137, 3358

\bibitem[{{McEvoy} {et~al.}(2015){McEvoy}, {Dufton}, {Evans}, {Kalari},
  {Markova}, {Sim{\'o}n-D{\'{\i}}az}, {Vink}, {Walborn}, {Crowther}, {de
  Koter}, {de Mink}, {Dunstall}, {H{\'e}nault-Brunet}, {Herrero}, {Langer},
  {Lennon}, {Ma{\'{\i}}z Apell{\'a}niz}, {Najarro}, {Puls}, {Sana},
  {Schneider}, \& {Taylor}}]{mce14}
{McEvoy}, C.~M., {Dufton}, P.~L., {Evans}, C.~J., {et~al.} 2015, \aap, 575, A70

\bibitem[{{Mokiem} {et~al.}(2007){Mokiem}, {de Koter}, {Vink}, {Puls}, {Evans},
  {Smartt}, {Crowther}, {Herrero}, {Langer}, {Lennon}, {Najarro}, \&
  {Villamariz}}]{mok07}
{Mokiem}, M.~R., {de Koter}, A., {Vink}, J.~S., {et~al.} 2007, \aap, 473, 603

\bibitem[{{Pasquini} {et~al.}(2002){Pasquini}, {Avila}, {Blecha}, {Cacciari},
  {Cayatte}, {Colless}, {Damiani}, {de Propris}, {Dekker}, {di Marcantonio},
  {Farrell}, {Gillingham}, {Guinouard}, {Hammer}, {Kaufer}, {Hill}, {Marteaud},
  {Modigliani}, {Mulas}, {North}, {Popovic}, {Rossetti}, {Royer}, {Santin},
  {Schmutzer}, {Simond}, {Vola}, {Waller}, \& {Zoccali}}]{pas02}
{Pasquini}, L., {Avila}, G., {Blecha}, A., {et~al.} 2002, The Messenger, 110, 1

\bibitem[{{Pecaut} \& {Mamajek}(2013)}]{pec13}
{Pecaut}, M.~J. \& {Mamajek}, E.~E. 2013, \apjs, 208, 9

\bibitem[{{Petermann} {et~al.}(2015){Petermann}, {Langer}, {Castro}, \&
  {Fossati}}]{pet15}
{Petermann}, I., {Langer}, N., {Castro}, N., \& {Fossati}, L. 2015, \aap, 584,
  A54

\bibitem[{{Pfuhl} {et~al.}(2014){Pfuhl}, {Alexander}, {Gillessen}, {Martins},
  {Genzel}, {Eisenhauer}, {Fritz}, \& {Ott}}]{pfu14}
{Pfuhl}, O., {Alexander}, T., {Gillessen}, S., {et~al.} 2014, \apj, 782, 101

\bibitem[{{Puls} {et~al.}(2008){Puls}, {Vink}, \& {Najarro}}]{pul08}
{Puls}, J., {Vink}, J.~S., \& {Najarro}, F. 2008, \aapr, 16, 209

\bibitem[{{Ram{\'{\i}}rez-Agudelo} {et~al.}(2013){Ram{\'{\i}}rez-Agudelo},
  {Sim{\'o}n-D{\'{\i}}az}, {Sana}, {de Koter}, {Sab{\'{\i}}n-Sanjul{\'{\i}}an},
  {de Mink}, {Dufton}, {Gr{\"a}fener}, {Evans}, {Herrero}, {Langer}, {Lennon},
  {Ma{\'{\i}}z Apell{\'a}niz}, {Markova}, {Najarro}, {Puls}, {Taylor}, \&
  {Vink}}]{ram13}
{Ram{\'{\i}}rez-Agudelo}, O.~H., {Sim{\'o}n-D{\'{\i}}az}, S., {Sana}, H.,
  {et~al.} 2013, \aap, 560, A29

\bibitem[{{Ryans} {et~al.}(2003){Ryans}, {Dufton}, {Mooney}, {Rolleston},
  {Keenan}, {Hubeny}, \& {Lanz}}]{rya03}
{Ryans}, R.~S.~I., {Dufton}, P.~L., {Mooney}, C.~J., {et~al.} 2003, \aap, 401,
  1119

\bibitem[{{Sabbi} {et~al.}(2013){Sabbi}, {Anderson}, {Lennon}, {van der Marel},
  {Aloisi}, {Boyer}, {Cignoni}, {de Marchi}, {de Mink}, {Evans}, {Gallagher},
  {Gordon}, {Gouliermis}, {Grebel}, {Koekemoer}, {Larsen}, {Panagia}, {Ryon},
  {Smith}, {Tosi}, \& {Zaritsky}}]{sab13}
{Sabbi}, E., {Anderson}, J., {Lennon}, D.~J., {et~al.} 2013, \aj, 146, 53

\bibitem[{{Sabbi} {et~al.}(2016){Sabbi}, {Lennon}, {Anderson}, {Cignoni}, {van
  der Marel}, {Zaritsky}, {De Marchi}, {Panagia}, {Gouliermis}, {Grebel},
  {Gallagher}, {Smith}, {Sana}, {Aloisi}, {Tosi}, {Evans}, {Arab}, {Boyer}, {de
  Mink}, {Gordon}, {Koekemoer}, {Larsen}, {Ryon}, \& {Zeidler}}]{sab16}
{Sabbi}, E., {Lennon}, D.~J., {Anderson}, J., {et~al.} 2016, \apjs, 222, 11

\bibitem[{{Sana} {et~al.}(2013){Sana}, {de Koter}, {de Mink}, {Dunstall},
  {Evans}, {H{\'e}nault-Brunet}, {Ma{\'{\i}}z Apell{\'a}niz},
  {Ram{\'{\i}}rez-Agudelo}, {Taylor}, {Walborn}, {Clark}, {Crowther},
  {Herrero}, {Gieles}, {Langer}, {Lennon}, \& {Vink}}]{san13}
{Sana}, H., {de Koter}, A., {de Mink}, S.~E., {et~al.} 2013, \aap, 550, A107

\bibitem[{{Sana} {et~al.}(2012){Sana}, {de Mink}, {de Koter}, {Langer},
  {Evans}, {Gieles}, {Gosset}, {Izzard}, {Le Bouquin}, \& {Schneider}}]{san12}
{Sana}, H., {de Mink}, S.~E., {de Koter}, A., {et~al.} 2012, Science, 337, 444

\bibitem[{{Schmitt}(1985)}]{sch85}
{Schmitt}, J.~H.~M.~M. 1985, \apj, 293, 178

\bibitem[{{Sim{\'o}n-D{\'{\i}}az}(2010)}]{sim10a}
{Sim{\'o}n-D{\'{\i}}az}, S. 2010, \aap, 510, A22

\bibitem[{{Sim{\'o}n-D{\'{\i}}az} {et~al.}(2017){Sim{\'o}n-D{\'{\i}}az},
  {Godart}, {Castro}, {Herrero}, {Aerts}, {Puls}, {Telting}, \&
  {Grassitelli}}]{sim17}
{Sim{\'o}n-D{\'{\i}}az}, S., {Godart}, M., {Castro}, N., {et~al.} 2017, \aap,
  597, A22

\bibitem[{{Sim{\'o}n-D{\'{\i}}az} \& {Herrero}(2007)}]{sim07}
{Sim{\'o}n-D{\'{\i}}az}, S. \& {Herrero}, A. 2007, \aap, 468, 1063

\bibitem[{{Sim{\'o}n-D{\'{\i}}az} \& {Herrero}(2014)}]{sim14}
{Sim{\'o}n-D{\'{\i}}az}, S. \& {Herrero}, A. 2014, \aap, 562, A135

\bibitem[{{Sim{\'o}n-D{\'{\i}}az} {et~al.}(2006){Sim{\'o}n-D{\'{\i}}az},
  {Herrero}, {Esteban}, \& {Najarro}}]{sim06}
{Sim{\'o}n-D{\'{\i}}az}, S., {Herrero}, A., {Esteban}, C., \& {Najarro}, F.
  2006, \aap, 448, 351

\bibitem[{{Sim{\'o}n-D{\'{\i}}az} {et~al.}(2010){Sim{\'o}n-D{\'{\i}}az},
  {Herrero}, {Uytterhoeven}, {Castro}, {Aerts}, \& {Puls}}]{sim10}
{Sim{\'o}n-D{\'{\i}}az}, S., {Herrero}, A., {Uytterhoeven}, K., {et~al.} 2010,
  \apjl, 720, L174

\bibitem[{{Trundle} {et~al.}(2007){Trundle}, {Dufton}, {Hunter}, {Evans},
  {Lennon}, {Smartt}, \& {Ryans}}]{tru07}
{Trundle}, C., {Dufton}, P.~L., {Hunter}, I., {et~al.} 2007, \aap, 471, 625

\bibitem[{{Uns{\"o}ld}(1942)}]{uns42}
{Uns{\"o}ld}, A. 1942, \zap, 21, 22

\bibitem[{{Walborn}(1972)}]{wal72}
{Walborn}, N.~R. 1972, \aj, 77, 312

\bibitem[{{Walborn} {et~al.}(2014){Walborn}, {Sana}, {Sim{\'o}n-D{\'{\i}}az},
  {Ma{\'{\i}}z Apell{\'a}niz}, {Taylor}, {Evans}, {Markova}, {Lennon}, \& {de
  Koter}}]{wal14}
{Walborn}, N.~R., {Sana}, H., {Sim{\'o}n-D{\'{\i}}az}, S., {et~al.} 2014, \aap,
  564, A40

\bibitem[{{Wegner}(1994)}]{weg94}
{Wegner}, W. 1994, \mnras, 270, 229

\end{thebibliography}

\begin{appendix}
\section{Estimates of projected rotational velocities}\label{vsini}
\subsection{Data preparation}

The data reduction for the individual exposures followed the
procedures discussed in Paper~I and \citet[][ hereinafter
  Paper~II]{duf12}. Because of the radial-velocity variations due to
binarity, care had to be taken when combining exposures. We have
undertaken simple numerical experiments to estimate the maximum range
in radial velocities (\dvr) of individual exposures that can be
combined before the estimation of the projected rotational velocity
becomes compromised. The spectrum from a main-sequence {\sc tlusty} model
\citep[with an effective temperature of 25\,000 K, logarithmic gravity
  of 4.0~dex and microturbulence of 5~\kms]{hub88, hub95, hub98,
  lan07, rya03, duf05} was convolved with an appropriate instrumental
profile. This was then convolved with a projected rotational
broadening function (with a given value of \vsini) and the resultant
spectrum shifted by a radial velocity \dvr\ before being combined with
the unshifted convolved spectrum. This simulates two observations of
the primary of the binary system with a velocity difference, \dvr.

The \ion{Si}{iii} line at 4552\AA\ in the combined spectrum was then
analysed using the Fourier-Transform methods discussed in Paper~II to
yield the `observed' \vsini. The results of these simulations are
summarized in Table~\ref{t_vsini_errors} for different choices of
\dvr\ and \vsini. As expected the `observed' \vsini\ estimates are
reliable for cases where either \dvr\ is small and/or \vsini\ is
large; these cases lie to the right of the dotted line drawn in the
Table. Note that as in reality we will be combining typically twelve
LR02 exposures that all lie within the radial-velocity range, \dvr,
this is a very stringent test and in reality the dotted line is likely
to lie further to the left (i.e. at lower values of \vsini).

Using these simulations, our analysis procedure was as
follows. Firstly all LR02 spectra were combined without any
wavelengths shifts using the procedures discussed by \citet{duf12} for
single stars. The projected rotational velocity, \vsini, was then
estimated and if it lay to the right of the dotted line, the estimate
was accepted. If it lay on or to the left of this line, the spectra
were recombined but shifted using the radial-velocity estimates of
\citet{dun15}; this procedure was appropriate for the metal line
spectra where no nebular emission was present. A further reduction was
also undertaken using spectra from only the best LR02 epoch (in terms
of signal-to-noise ratio) plus any other epochs with similar radial
velocities (i.e. leading to a range of radial velocities, $\dvr \leq
30$~\kms); this was preferable for the \ion{He}{i} features which are
affected by nebular emission (see Papers I and II for more details).

\begin{table}

\caption{Simulation of the effects of radial velocity variations on the estimation of the projected rotational velocity from the \ion{Si}{iii} line at 4552\AA. Dashes indicate that the profile appear double peaked and no attempt to estimate a projected rotational velocity was attempted. The region where the estimates should be reliable is at the upper right and is delineated by dotted lines. All values are in \kms.}\label{t_vsini_errors}

\begin{center}

\begin{tabular}{crrrrrr}

\hline\hline

\vspace*{-0.25cm}\\
\dvr  & \multicolumn{5}{c}{Stellar \vsini}   \\
&40 & 80 & 120 & 160 & 200 \\

\hline
30    & \multicolumn{1}{;{2pt/1pt}r}{40}  & 80   & 119 & 159 & 199 \\\cdashline{2-2}[2pt/1pt] 
40    & 51  & \multicolumn{1}{;{2pt/1pt}r}{79}   & 121 & 160 & 200 \\
60   & 76   & \multicolumn{1}{;{2pt/1pt}r}{79}   & 121 & 161 & 200 \\\cdashline{3-3}[2pt/1pt]
80   & -      & 101   &  \multicolumn{1}{;{2pt/1pt}r}{119} & 162 & 200 \\
100 & -      & 132  & \multicolumn{1}{;{2pt/1pt}r}{124} & 162 & 199 \\\cdashline{4-4}[2pt/1pt]
120 & -      & -       & 154 & \multicolumn{1}{;{2pt/1pt}r}{161} & 203 \\
\hline
\end{tabular}
\end{center}

\end{table}

\subsection{Methodology for estimating projected rotational velocities}\label{s_vsini}

The procedures for estimating projecting rotational velocities for the
B-type binary sample were very similar to those undertaken by
Paper~II. In summary, a Fourier-Transform (FT) methodology \citep{sim07}
was employed. This has been used to estimate projected rotational
velocities in VFTS main-sequence stars \citep[see,][and
  Paper~II]{ram13} and supergiants \citep{mce14} and has also been
widely used to study the different mechanisms contributing to the
broadening of spectral lines in  early-type stars
\citep[see, for example][]{dufsmc06, lef07, mar07, sim10, fra10, sim14, sim17}.

As discussed by  \citet{sim07} the FT method  should be able to explicitly separate the rotational broadening from other broadening such as natural, instrumental and natural. The approach relies on the convolution theorem \citep{gra05}, viz. that the Fourier transform of convolved functions is proportional to the product of their individual Fourier Transforms. It then identifies the first minimum in the Fourier transform for a spectral line, which is assumed to be the first zero in the Fourier transform of the rotational broadening profile with the other broadening mechanisms exhibiting either no minima or only minima at higher frequencies. Further details on the implementation of this methodology can be found in \citet{sim07} and paper II.

These estimates were supplemented by
fitting rotational broadened profiles directly to the
observed spectra. As this approach neglects the contribution of the
intrinsic line profile and the instrumental profile, it
effectively provides an upper limit (see Paper~II for a detailed
discussion and comparison of the different approaches) and is
used here solely as a consistency check.

Two sets of absorption lines were used, which were identical to those
adopted in Paper~II. For targets with relatively low projected
rotational velocities ($\vsini \la 150$~\kms), metal and non-diffuse
\ion{He}{i} were adopted; for other targets, both non-diffuse and
diffuse \ion{He}{i} lines were used. Further details and comparison of
results estimated from the two sets of lines can be found in Paper~II.

In Tables~2 and~3 (only available online), the estimates from the individual lines are listed for all the non-supergiant targets that are believed to be the primaries of binary systems -- these Tables have an identical format
to those of Tables~3 and~4 of Paper~II, where again further details
can be found\footnote{Projected rotational velocities are also listed for 16 additional targets that were excluded from Paper~II as at that time they were believed to be binaries but have subsequently been designated as 'single'. Additionally estimates for 7 targets that are now designated as binaries have been given previously in Paper~II.}.
\end{appendix}

\end{document}